\documentclass[12pt]{article}

 \usepackage{amsfonts}
 \usepackage{amssymb,amsmath}
 \usepackage{graphicx} \usepackage{epstopdf}
 \newcommand{\crlb}[1]{\label{#1}\\[2pt]}
 
 \newcommand{\crld}[1]{\label{#1}}
 \newcommand{\eela}[1]{\quad\hbox{\scriptsize{#1}}\label{#1}\end{eqnarray}}
 \newcommand{\eelb}[1]{\label{#1}\end{eqnarray}}
 
 \newcommand{\newsecb}[2]{\section{#1}\label{#2}\setcounter{equation}{0}}

 \newcommand{\nolabels} {\def\eel{\eelb}\def\eeql{\eeqlb}  \def\crl{\crlb} \def\newsecl{\newsecb}\def\bibiteml{\bibitem} \def\citel{\cite}\def\labell{\crld}}
\newcommand{\eeqla}[1]{\quad\hbox{\scriptsize{#1}}\label{#1}\end{aligned}\end{equation}}
\newcommand{\eeqlb}[1]{\label{#1}\end{aligned}\end{equation}}

\newcommand\publishversion{\nolabels\setlength{\textheight}{8.3in}\setlength{\oddsidemargin}{0in}
   	 \setlength{\textwidth}{6.3in}\setlength{\topmargin}{-0.2in}}
\def\beq{\begin{equation}\begin{aligned}}		\def\eeq{\end{aligned}\end{equation}}
\def\be{\begin{eqnarray}}  					\def\ee{\end{eqnarray}}		%\be and \ee will become obsolete in due time.
\def\bi#1{\begin{itemize}\item[#1]} 			\def\itm#1{\item[#1]} 			\def\ei{\end{itemize}} 
  \def\eqn#1{(\ref{#1})}
   	 \def\fn{\footnote}	  		 

		 \def\a{\alpha}   \def\b{\beta}   \def\d{\delta}   \def\k{\kappa}       
		       \def\r{\varrho}      \def\t{\tau}    
    		\def\G{\Gamma}  	\def\D{\Delta}   \def\e{\varepsilon} 
\def\m{\mu}	    		\def\f{\phi}        		     		\def\vv{\varphi}     \def\n{\nu}
 	 		\def\s{\sigma}     	\def\tht{\theta}      	 
	     		          		              	\def\w{\omega}  
\def\W{\Omega}    		  		\def\dd{{\rm d}} 		\def\HH{{\mathcal H}}  
\def\OO{{\mathcal O}} 		     
 		      
\def\pa{\partial}			\def\ra{\rightarrow}	
\def\bra{\langle} 		\def\ket{\rangle}

	\def\iss{\ =\ }
\def\bal{$\bullet$} 
\def\fract#1#2{{\textstyle\frac{#1}{#2}}}	 	 \def\fractje#1#2{{\scriptstyle\frac{#1}{#2}}}	
\def\ffract#1#2{\raise .2 em\hbox{$\scriptstyle#1$}\kern-.3em/\kern-.2em\lower .15 em \hbox{$\scriptstyle#2$}}
\def\half{\fract12}					\def\halfje{\fractje12}
\def\tl#1{\tilde{#1}} 
\def\ex#1{e^{\textstyle#1}} 		\def\qqquad{\qquad\qquad}		\def\qqqquad{\qqquad\qqquad}
\def\bpmatrix{\begin{pmatrix}} 			\def\epmatrix{\end{pmatrix}}
\def\bmatrix{\begin{matrix}} 			\def\ematrix{\end{matrix}} 
\def\bcenter{\begin{center}}			\def\ecenter{\end{center}}

\def\lowerheightfig#1#2#3{\(\raise-#1\hbox{\includegraphics[height=#2]{#3}}\)}
\def\lowerwidthfig#1#2#3{\(\raise-#1\hbox{\includegraphics[width=#2]{#3}}\)}
\def\th{\({}^{\mathrm{th}}\)}		 \def\vecc#1{\hskip.08em\vec{#1}\,}
\publishversion
\begin{document}

\begin{titlepage}
 \title{\vskip 20mm \LARGE\bf The firewall transformation for black holes \\  and some of its implications}
\author{Gerard 't~Hooft}
\date{\normalsize Institute for Theoretical Physics \\ Utrecht University  \\[10pt]
 Postbox 80.089 \\ 3508 TB Utrecht, the Netherlands  \\[10pt]
e-mail:  g.thooft@uu.nl \\ internet: 
http://www.staff.science.uu.nl/\~{}hooft101/}
 \maketitle

\begin{quotation} \noindent {\large\bf Abstract } \medskip \\
A promising strategy for better understanding space and time at the Planck scale, is outlined and further pursued. It is explained in detail, how black hole unitarity demands the existence of transformations that can remove firewalls. This must then be combined with a continuity condition on the horizon, with antipodal identification as an inevitable consequence. The antipodal identification comes with a $CPT$ inversion. 

We claim to have arrived at `new physics', but rather than string theory, our `new physics' concerns new constraints on the topology and the boundary conditions of general coordinate transformations. 

The resulting theory is conceptually quite non trivial, and more analysis is needed. A strong entanglement between Hawking particles at opposite sides of the black hole is suspected, but questions remain.
A few misconceptions concerning black holes, originating from older investigations, are discussed. \\[10pt]
\end{quotation}
% Version April, 2017 \hfill  Typeset \today. \\
 %{\small (contains only minor editorial improvements compared to the december 2016 paper (v1))}

\end{titlepage}

\eject
\setcounter{page}{2}

\def\inn{{\mathrm{in}}}  \def\outt{{\mathrm{out}}} 
\def\Pl{{\mathrm{Planck}}}

% \tableofcontents 

\newsecl{Introduction}{intro}

After at least 4 decades of intense studies all over the world, it has become evident that the dynamical laws of space, time and matter will have to be reformulated at time and distance scales comparable to the Planck scale,\fn{No assumptions are made concerning extra dimensions\,\cite{extradim}. These would rearrange dynamical variables in such a way that they each seem to occupy much more space in the physical dimensions. This would lead to effective ultimate scales of physics different from, and less exotic than, \eqn{Planckunits}. Our basic conclusions will however not be affected.}\,\fn{In contrast, the Planck \emph{momentum} has the remarkably mundane value of just over  \(6.5\,\)kg\,m/s\ .}
\beq L_\Pl&=1.6162\times 10^{-33}\,\mathrm{cm}\ ,&\qquad T_\Pl&=5.391\times 10^{-44}\,\mathrm{s}\ ,\\
M_\Pl&=21.765\,\m\mathrm{g}\ ,&E_\Pl&=1.2209\times 10^{28}\,\mathrm{eV}\,. \eeql{Planckunits}
However, in spite of admirable advances in superstring theories\,\cite{sustr}, and competing approaches such as loop quantum gravity\,\cite{loopgrav}, we still do not have completely consistent models that elucidate Nature's book keeping system\,\cite{bookkeeper} at this scale: what are the physical degrees of freedom, how are they arranged in space, how do they evolve in time, and to what extent are constraints on locality, unitarity, positivity, stability and finiteness obeyed? Which symmetries do we have, which of these are exactly valid, and which symmetries are spontaneously or explicitly broken? How should we formulate the boundary conditions? And so on. As we shall show, there are still surprises to be expected.

The fact that our modern approaches fail was demonstrated embarrassingly clearly when it was realised that there is a \emph{firewall problem}\,\cite{firewall} in black hole physics. Actually, this firewall problem was just one way of phrasing the information paradoxes in black holes\,\cite{Polch}, and it demonstrates, once again, that some fundamental physical principles must hold that are not understood at all by many researchers today.

Black holes are indeed the most manifest structures where our present understanding is seen to be hopelessly inadequate. As was emphasised by the present author at several occasions before\,\cite{GtHBH, GtHSmatrix}, the best way to make further progress is therefore to address black holes up front, demand laws of physics that guarantee logically comprehensible and consistent behaviour of these structures, and inspect to what extent our present formulations will have to be adjusted or sharpened.

The earliest ideas about the states black holes can be in, as deduced from Hawking's observation that black holes emit particles with what looks like a thermal spectrum, were that black holes cannot be in pure quantum states, but will always be in mixed states, such as what we have in thermal equilibrium. Requiring maximal accuracy in formulating their evolution laws, however, forces us to search for formalisms in terms of pure quantum states. This does come at a price to be paid. In the present work, we show what is required to arrive at a description of  black holes in terms of pure quantum states. The price is new physics, as was emphasised earlier\,\cite{GtHBH}--\cite{GtHrecent2}.

There is a number of points that we should keep in mind. One is that wild guesses concerning possible answers, such as `novel uncertainty relations', will be almost fruitless, as history shows. The best thing to do is to split our problems into small pieces, and try to address each of these small fragments of questions in turn. Every now and then, such fragmented questions will lead to surprises.  It helps enormously if we can convince ourselves of the correctness of our partial answers, and it is these that we should be able to use as new starting points for our next steps.

In this paper, we shall primarily make use of a partial answer that we claim to have arrived at recently\,\cite{GtHrecent2}: the necessity of revising the \emph{boundary conditions} for Nature's degrees of freedom at the horizon of a black hole, summarised in Appendix \ref{KSantipodes}. Since our analysis started out with our desire for consistent descriptions of \emph{stationary} (or approximately stationary) black holes, it was not immediately clear how the revised boundary conditions should have been enforced during the {formation} of a black hole, but, in a somewhat formal fashion, one may well argue that, during black hole formation,  the horizon starts out stretching over an infinitesimally tiny region; it opens up at a single point\fn{If the collapse was in a spherical shell of matter, this point lies well inside the shell, surrounded by a local vacuum. This makes it easy to study. One then might conclude that the horizon first forms on a \emph{fractal} subspace of space-time, but since the scale at which this fractal extends may end up to be small even in Planck units, we ignore this complication in this paper.} in space and time.  At that single point, it now appears to be necessary to revise the structure of this infinitesimal horizon to obey the new boundary condition, but since all this should happen at Planckian dimensions, the revision needed in our laws of Nature here can easily be argued to have escaped our notice up to today; it may be seen as merely a mild, point-like singularity at Planckian dimensions\fn{During a black hole's final evaporation event, the topologically non-trivial space-like features should disappear the same way as they came.}. We elaborate this further in Section \ref{collapse}.

After the horizon opens up, a black hole can grow quite big;  the black hole horizon area grows rapidly towards macroscopic sizes during collapse,  and as our modified boundary condition keeps track, it turns space and time into a non-trivial topological manifold.  
Our new boundary condition must be locally imperceptible, but its implications will be sizeable. As a starting point we may suspect that the entire process of the formation, evolution, and the final explosion of a black hole may be seen as an \emph{instanton event}, an instanton that has a trivial Minkowskian boundary yet it is locally non-trivial in a way that we shall explain (Section~\ref{instanton}).

We emphasise that,  nevertheless, our modified boundary condition will not affect the visible properties of a black hole in the classical limit.  Also,  we shall ensure that the modified boundary condition is of a kind that is not directly observable for a \emph{local} observer, that is, an observer who can only see his/her immediate environment. So what we call `new physics' is still completely in line with ordinary quantum mechanics and general relativity.

The boundary condition that we shall arrive at is characterised as an \emph{antipodal identification}. In short, what it means is that the region of space-time inside the horizon is removed completely, as if by surgery, after which the edges are glued together by identifying the antipodes. This is continued throughout the lifetime of the black hole.\fn{Do keep in mind that, strictly speaking, the horizon is entirely timeless.} It is important, subsequently to insist that, locally, space and time remain smooth across the seams, while particles, including the information they carry,  can cross. The seams must be locally invisible---only \emph{global} observers notice this boundary condition. We argue that the antipodal mapping is the only way to attach the edges together such that strict geometrical conditions are obeyed. 

It boils down to a single ``new physics" ingredient in black hole physics as soon as quantum effects are being considered. To explain this, let us define the asymptotic region of space-time as the region \(|r|\ra\infty\), which splits into five parts:\\
 \(\infty^{+}\), where in natural units \(t\gg|r|\gg M\),\\
  space-like infinity \(\infty^0\) with \(|r|\gg|t|\),\\
  the region \(\infty^{-}\), where \(-t\gg|r|\gg M\),\\
   and finally the two light cone regions \(\mathcal{J}^\pm\) separating the previous three domains. \\
   Then we impose:
\begin{quote} 
\emph{When fields on a manifold are quantised, it is essential that the entire asymptotic domain of the manifold maps one-to-one onto that of ordinary space-time, while preserving the metric. It must be possible to find time-like paths that connect all space-time points in \(\infty^+\)  to all points in \(\infty^-\) }
 \end{quote}
This condition is not obeyed by the standard, continuous extension of the classical Schwarzschild metric: every space-time point in the physically observable part of the universe it describes is mapped onto \emph{two} points in the Kruskal-Szekeres coordinates. These two points are always space-like separated from one another. If we would allow this situation to describe a black hole, we would end up with two universes connected by a worm hole, as is well-known. These two universes would communicate to one another quantum mechanically (that is, they are entangled), which causes the well-known violation of unitarity. We find that this difficulty is completely resolved by identifying points in region \(I\) by their antipodes in region \(II\). It is a folding, which avoids any singularity (cusp-like or otherwise). In contrast, such a folding, to be referred to as \emph{antipodal identification},  would not be possible in \emph{flat} space-time without cusp singularities, see Appendix \ref{KSantipodes}. Note that the \(r\ra 0\) singularities in Schwarzschild black holes, as well as the inner horizon in Kerr and Reissner Nordstr\"om black holes, occur in a space-time region that is entirely avoided in our treatment.\fn{We acknowledge the observation of a referee that the above was not formulated accurately in a previous version of this paper, so that confusions could arise; presumably more accurate discussions are possible in a mathematical language that was avoided here.}

As will be demonstrated (section \ref{qustates}), mapping the Schwarzschild metric onto the space-time metric of a local observer forces us to glue together regions in such a way that \emph{time-inversion} takes place. Inverting the time direction is associated with an interchange of creation operators and annihilation operators,\fn{In an earlier version of this paper it was concluded that, therefore, at the horizon we must glue the vacuum state onto a ``completely full state", but this does not seem to be necessary; we do perturbative quantum field theory near the vacuum state in all regions of the Penrose diagram. See the discussion in Section \ref{qustates} and Fig.~\ref{flippingregions.fig} there.} in the sense of Bogolyubov transformations.\fn{Provided care is taken to maintain unitarity, see Footnote~\ref{aadagger} in Section \ref{microstates}.} 

We shall insist that we begin by limiting ourselves to \emph{soft particles}, which are defined as particles whose gravitational fields are either negligible or sufficiently weak to allow for a description in terms of \emph{perturbative} gravity. This will be further explained in Section~\ref{qustates}. We are dealing with an essential and highly non-trivial demand here, since the time evolution may seem to turn soft particles into hard particles -- particles which do have sizeable effects on the curvature of space and time. These particles will have to be transformed into something else, as we shall see.
The gravitational fields of these particles, in particular the \emph{hard} particles, will be taken care of in due time. Thus, the fact that we allow ourselves to have soft particles only in our Penrose diagram will be justified a posteriori.

Crossing the horizon from a given point to its antipode will be associated with such a time inversion, and as such also interchanges creation and annihilation operators. This allows the embedding of a \emph{time-like M\"obius strip} in our space-time, and it has the remarkable effect that the Hartle-Hawking state links positive energy particles at the horizon with antiparticles at the antipodes, which again have positive energies, resulting in entanglement between positive energy particles only, see Chapter~\ref{Hawk2}.

Since quantum field theories are \(CPT\) invariant rather than just \(T\)-invariant, we expect that the different domains adjacent to a horizon in the Penrose diagram will be visible to the outside observer through \(CPT\) inversions. Because of the \(CPT\) inversion, the global, causal arrow of time no longer coincides with the local arrow of time. This implies a departure from earlier ideas expressed by this author, called black hole complementarity, where it was assumed that causal ordering should be kept untouched. Black hole complementarity addressed the interior regions of the black hole as representing particles emerging later, while in our new description, the black hole simply has no interior at all. When putting everything together, one finds that this latter formalism is far more satisfactory: \emph{nothing ever escapes to the interior region of the black hole, since there is no interior region.}

At first sight, it may seem that our way of handling space and time near a black hole, will make a decent quantum field theoretic description of the elementary particles hopelessly inadequate. However, as it turns out, the opposite is true: our apparently drastic rearrangement of the space-time continuum is exactly what is needed to arrive at pure quantum states for the black hole, and to obtain a unitary scattering matrix, so as to eradicate both the black hole information problem and the firewall problem, while meticulously respecting the laws of general relativity. 

As our work is still in progress, there are numerous issues still remaining; we discuss some of these as representing new challenges. Together, they constitute a new and systematic strategy for future investigations. 

We emphasise that, barring possible minor mistakes,\fn{This paper had to be rewritten several times because of small mistakes in earlier versions. Perhaps this is to be blamed to the fact that it has only one author.} our conclusions are solid and inevitable, comparable to the much more grandiose introduction of the theory of general relativity itself, which may be seen as an inevitable description of the gravitational force if one relies on special relativity at small scales and the equivalence of gravitational and inertial masses.

We also emphasise the importance of applying our modified rules in the case of the quantum black hole. At first sight, the replacement of the analytic continuation of Schwarzschild space-time by the one obtained by identifying antipodal points with region \(II\) of the Penrose diagram, may seem to be a minor modification, but it has big effects. Formerly, the Hartle-Hawking vacuum used to be applied in such a way that observers do not have access to the hidden sector, so that the resulting state become a thermodynamically mixed state. In our description, this state remains a pure state for the outside observer; the part that used to be hidden actually describes the other side of the same black hole. Now that we have only pure states, the resolution of the ``information problem" is straightforward.

Our rules remove the `insides' of a black hole.
\newsecl{Black hole Penrose diagrams}{Penrose}		

 \setcounter{footnote}{0}

\begin{figure}[h!]
	\includegraphics[width=440pt]{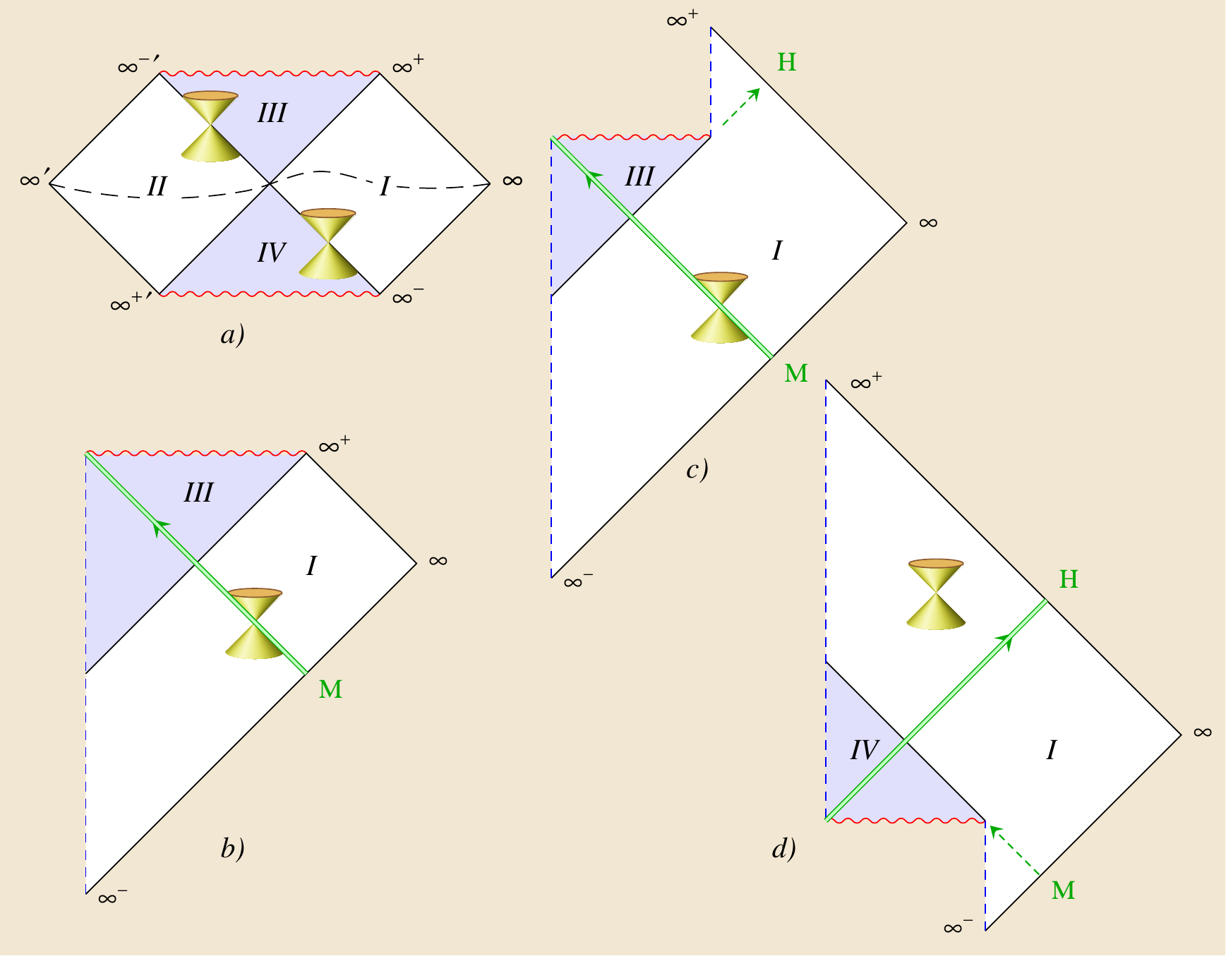} \\[-30pt] 
	\begin{quote}
	\begin{caption}{ }{\small{ Penrose diagrams for Schwarzschild black holes. $a)$ The ``eternal" black hole, showing regions \(I,\ II,\ III\), and \(IV\). Dotted line: Cauchy surface (see text). $b)$ A black hole originating from imploding matter. Matter \(M\) causes curvature along the diagonal shown, rendering  \(II\) and \(IV\) invisible. Dotted line: $r=0$. $c)$ Including the effect of Hawking radiation $(H)$ making the metric regular at \(t\ra\infty\), according to Hawking. $d)$ The \(CPT\) image of situation $c$. Imploding matter and Hawking matter interchange places. Local light cones everywhere are under \(45^\circ\) (cones shown).}}
	\labell{penroses.fig}
	\end{caption}\end{quote}
\end{figure}

The Penrose diagram\cite{penrose} of a space-time metric is obtained by choosing light cone coordinates \(u^+\) and \(u^-\), such that \(g_{++}=0\) and \(g_{--}=0\), implying that these coordinates are tangent to the local light cones. For black holes, the remaining two coordinates are the angles \(\tht\) and \(\vv\). Picturing the light cone  coordinates as tilted by \(45^\circ\), one always gets space-like coordinates running horizontally and time-like ones vertically, while the local light cones\fn{In the \(\tht,\vv\) directions, the light cones may be more complicated; in the unphysical regions of the Kerr and Kerr-Newman black holes,  closed time-like curves may emerge. As these are located far beyond the horizons, there will be no need to consider them here.} are oriented everywhere as pictured in Figure~\ref{penroses.fig}. In a Penrose diagram, all time-like geodesics with constant angular coordinates \(\tht\) and \(\vv\), go in a vertical direction, more steeply than \(45^\circ\).
Since one can keep these properties unaltered when the coordinates \(u^\pm\) are arranged to occupy compact domains, one can compress the entire universe in compact Penrose diagrams. Thus, the location of time-like infinity \((\infty^\pm)\) and space-like infinity \((\infty)\) can be indicated in the diagram, see Fig.~\ref{penroses.fig}.

If all effects of matter that may have caused the black hole to form, in the near or distant past, as well as matter originating from Hawking radiation, are ignored, one gets the maximally extended Schwarzschild solution, Fig.~\ref{penroses.fig}$a$. By inspecting how information spreads in such a universe, one then finds four distinct regions, labelled as \(I,\ II,\ III\) and \(IV\). Only regions \(I\) and \(II\) are connected to asymptotic space-like and time-like infinity, but they are only connected to each other through a two-dimensional surface, the origin of the Penrose diagram; they have \emph{different} asymptotic regions, indicated as \(\infty,\ \infty^\pm\) in region \(I\), and \(\infty'\) and \({\infty^\pm}'\) in region \(II\).

In the standard picture, nothing more is imposed such as antipodal identification; regions \(I\) and \(II\) could describe either entirely different universes, or perhaps different parts of the same universe, separated in space- and/or time-like directions, by as many light years as one can imagine. In contrast, regions \(II\) and \(IV\) are absent in Fig.~\ref{penroses.fig}$b$ and $c$. There, we see that they are replaced by region \(III\), so that it is often concluded that regions \(II\) and \(III\) actually describe the \emph{insides} of the black hole, whatever it may be that the insides of a black hole may look like. This is probably wrong, as we shall see. We return to the relation between regions \(I\) and \(II\) later in this section.
						%%%%% ---- %%%%%
						
First, we must dwell on an other urgent question. What exactly should it mean to distinguish `eternal' black holes from black holes that were once formed by gravitational collapse of ordinary matter? In \emph{classical} black holes, that is, black holes where quantum mechanical effects are assumed to be insignificant (as would be normal practice in standard general relativity), such distinctions would be unnecessary, or even meaningless. The differences between Figures \ref{penroses.fig}$a$ and $b$, are only in their distant past, but not in the present or the future. All no-hair theorems point to the verdict that black holes will all look and behave identically, apart from three parameters, being mass, charge and angular momentum.\,\cite{nohair}

Yet, when quantum mechanics comes into play, things do seem to be different. Since quantum mechanics strongly indicates that black holes emit particles, the future evolution of the Penrose diagram cannot quite be as in Figure \ref{penroses.fig}$a$ or $b$, but rather something like Fig.\ \ref{penroses.fig}$c$, as was advocated by Hawking\,\cite{HawkingPen}. There is a problem with that as well, however. If we wish to describe \emph{all} quantum states a black hole can be in, then surely one should expect that the symmetries of the system before quantisation, should also be reflected by the quantum system. This is not the case for Fig.~\ref{penroses.fig}$c$. Why is it not symmetric under time reversal? General relativity and quantum mechanics both are. One must conclude that we have not yet seen all possible Penrose diagrams; we should also brace ourselves for diagram \ref{penroses.fig}$d$.

How all these different possible Penrose diagrams can play their roles for black holes is one of the subjects of this paper. In a sense, they may all be right, but then, the theory claiming that they must represent the truth faithfully, cannot be correct. Indeed, we shall propose an important modification of how one should look at black holes. In this work, we make use of new transformations showing that all diagrams in Fig.~\ref{penroses.fig} can be used (see Section \ref{clones}). One then finds that the maximal extension of the Schwarzschild metric, shown in Fig.~\ref{penroses.fig}$a$, is the most useful one. It exhibits perfect symmetries under the exchange of regions \(I\) and \(II\) and/or regions \(III\) and \(IV\). These symmetries are not only special for the Schwarzschild metric, they also hold for the Reissner-Nordstr\"om, Kerr and the Kerr-Newman metrics. The most important reason for us to insist on using the eternal diagram of Fig.~\ref{penroses.fig}$a$, is that only that diagram allows us to use the analytic extension towards a domain describing all parts that we need of the neighbourhood of the region where future and past event horizons cross. This is exactly what is needed to apply the powerful laws of General Relativity right there.

In summary, the usual distinctions made between the Penrose diagrams $a$--$d$ in Fig.~\ref{penroses.fig} can no longer be maintained when we do quantum mechanics. This is because, if we want to consider Schr\"odinger equations to be applied at time \(t\approx 0\), we need the general wave functions of all particles in the present, so that (in the Heisenberg picture) the far past, as well as the far future, will consist of superpositions of all possible states. The space-time metric of the diagrams in Fig.~\ref{penroses.fig}$b$--$d$, requires the momentum distributions at very early and very late times to be precisely known, while the wave functions used at \(t\approx 0\) will be too general superpositions of states. In particular, if we want to consider a black hole in a pure quantum state such as the Hartle-Hawking state at \(t\approx 0\), its initial state, at \(\hbox{$t\ra-\OO(M^3)$}\) in Planck units, cannot be described as a classically collapsing object. Similarly, since at \(t\ra+\OO(M^3)\), when the black hole explodes, its standard description would be in terms of a density matrix; in our work we would treat the Hartle-Hawking state as a pure quantum state, but this would have to be replaced by quantum superpositions of explosions at different times. Due to such difficulties, which are comparable to the measurement problem in ordinary quantum mechanics,  we shall find that classical Penrose diagrams are unsuitable for describing pure quantum states at times \(t\) when \(|t|\gg M\log M\) in Planck units. We shall establish new procedures to justify the use of the Penrose diagram for an eternal black hole (Fig.~\ref{penroses.fig}$a$), with only soft particles added, being in a pure but generic quantum state while \(|t|\ll M\log M\).

Now we can return to our promise to discuss the link between region \(I\) and region \(II\). In this paper, we shall explain why antipodal points of the horizon have to be identified, with the consequence that regions \(I\) and \(II\) refer to opposite hemispheres of one and the same black hole.\fn{The possibility to have such an identification was first mentioned by Sanchez and Whiting\,\cite{SanchezWhiting} as far as the author is aware.} In the entire spacetime domain covered by both \(I\) and \(II\), as well as their close neighbourhoods, we have 
	\be r\ \gtrapprox\ 2GM\ . \eel{rgthM}
This is why it will never happen that two \label{pagecensornote} spacetime points that are much closer together than \(2GM\), are postulated to be identified.\fn{It does happen at the \emph{singularity} (the wavy lines in Fig.~\ref{penroses.fig}), but the singularity is way beyond the infinite future for the distant observer, so it will not be given any physical significance in this paper; the fact that the singularity is far beyond the infinite future makes it totally harmless in practice. This so-called \emph{cosmic censorship} is often not understood by laymen mystified by black holes (providing references here would be too embarrassing).\label{censornote}} This implies that local space-time continuity is not affected by the antipodal identification on the black hole horizon---there \emph{would} have been such a problem if we would have tried an antipodal identification in locally flat Minkowski space-time; in ordinary polar coordinates, a conical singularity would arise near \(r\approx 0\). This is why antipodal identification can \emph{only} be considered as a harmless (yet important) topological modification if applied to exotic situations such as these occur in the geometry of black holes.

The most delicate part of our new theory, however, is not the fact that we identify antipodal points, and not even the fact that we transform away the bad or unwanted regions of the Penrose diagram; it is the way in which particles are expected to transmit the information they carry, across the horizon. Since this information is transmitted entirely because of the gravitational back reaction, it is wrong to ignore that back reaction altogether, as is often done.

\newsecl{The horizon}{horizon}
\begin{figure}[h!]
	\qqqquad \includegraphics[width=200pt]{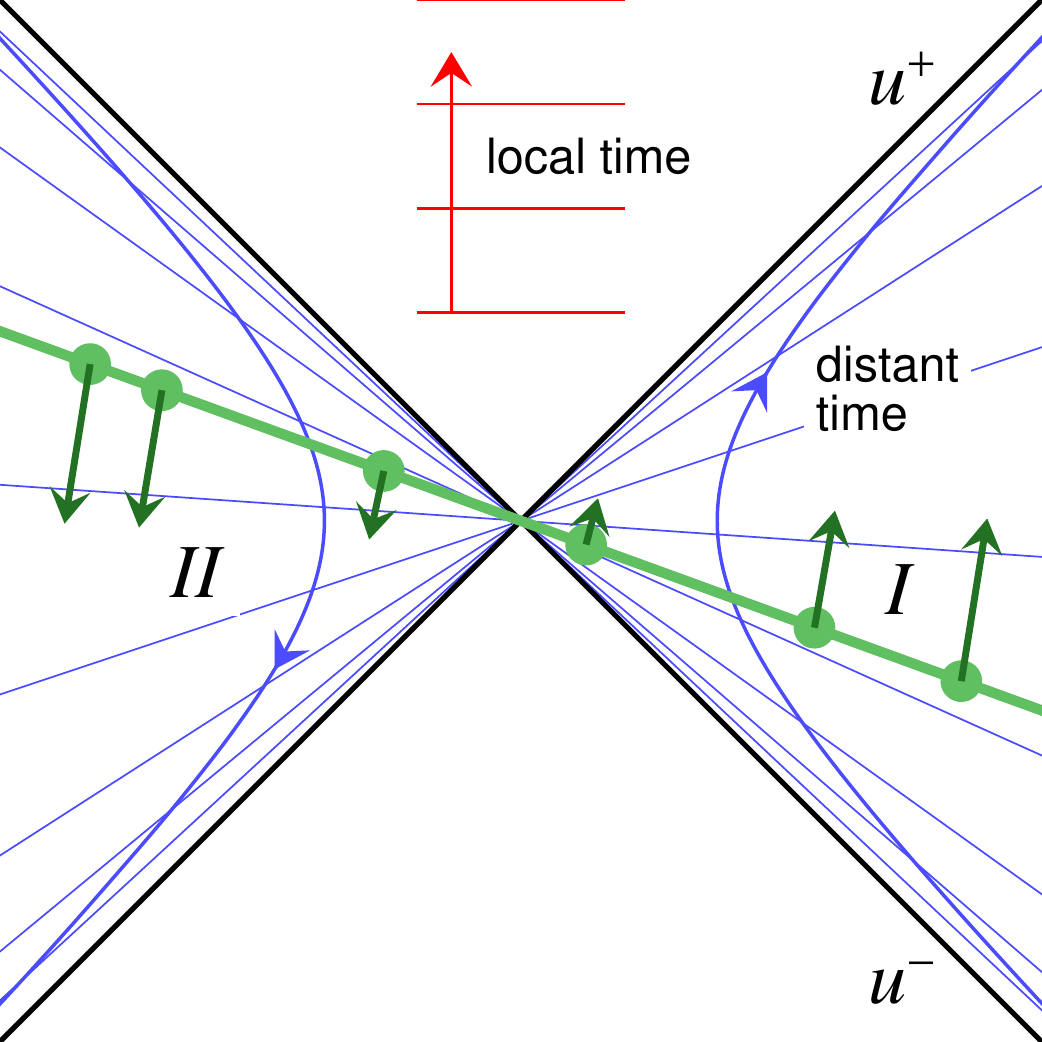} \begin{quote}
	\begin{caption}{ }{\small{Local light cone coordinates \(u^\pm\) near the horizon. The local time coordinate points upward everywhere, but the time coordinate for distant observers goes up in region \(I\) and down in region \(II\). Also shown is a local Cauchy surface. As the distant time variable proceeds, particles on this Cauchy surface move upwards in \(I\) and downwards in \(II\) (arrows).}}
	\labell{localhorizon.fig}
	\end{caption}\end{quote}
\end{figure}

In the Penrose diagrams of Fig.~\ref{penroses.fig}, it is often assumed that the time coordinate proceeds forward (in this paper usually indicated as upward) everywhere.  Indeed, this is what one would expect if region \(II\) were assumed to describe `the inside' of a black hole. In our present picture, however, it will be seen to be mandatory to switch the direction of time just in accordance with the global Schwarzschild time parameter \(t\), also when the quantum micro-states are discussed. This is because we wish to describe how the black hole evolves as seen by a distant observer. Thus the evolution process is postulated as indicated in Fig.~\ref{localhorizon.fig}.

In a local observer's description, a Cauchy surface stretching all across the Penrose diagram, can be postulated to contain only pure quantum states of elementary particles (the dots in Fig.~\ref{localhorizon.fig}). 
\emph{If we omit the mutual gravitational interaction between these particles}, general relativity tells us precisely how the evolution of these particles proceeds; in particular, they stay inside their regions \(I\) and \(II\). We can use the Standard Model to describe how their fields evolve.  Note, at this point, that, as seen by a local observer, the particles in region \(II\) will evolve backwards in time. We shall insist that the evolution will be forwards in time as seen by distant observers.

Consider a particle with mass \(\m\), close to both horizons, where we ignore local curvature of the metric. Let its  momentum in light cone coordinates be \(p=(p^+,p^-,\tl p)\), where \(\tl p\) is the transverse component. On mass shell, we have
\be 2p^+p^-+\tl p^2+\m^2=0\ .\eel{massshell}
The (non-gravitational part of) the evolution law is that time \(t\) generates Lorentz transformations. Defining the scaled time \(\t=t/4GM_{BH}\), we have
\be \tl p\ \hbox{ and }\ m\ \hbox{ stay constant, }\ p^-(\t)=p^-(0)e^\t\ ,\ \ p^+(\t)=p^+(0)e^{-\t}\ . \eel{evolv} 

The gravitational interaction between particles of matter is much less trivial, however; it will have the effect of shifting particles around in  directions along this Cauchy surface, as we shall see, allowing them to cross the horizon without much ado. It will be clear that this observation will be of crucial importance. In particular, we must formulate our theory for the entire Cauchy surface, stretching over both regions \(I\) and \(II\) (see dotted line in Fig.~\ref{penroses.fig}$a$).
\def\full{{\mathrm{full}}}

\subsection{Quantum states. Regions $I$ and  $II$. In- and out-particles. Soft and hard particles.\labell{qustates}}

However, there is the apparent complication mentioned at the end of Section~\ref{intro}: in region \(II\), time, as seen by a distant observer, runs backwards. Particles, running backwards in time, will be associated to quantum wave functions evolving as \(e^{+iEt}\) instead of \(e^{-iEt}\), and hence carry negative energies as seen by a distant observer. 
Nevertheless, the distant observer will need to describe his world in terms of positive energy particles. What happens, as it will turn out, is that, in region \(II\), the energy as seen by the distant observer is minus the energy experienced by an `observer' who would look at region \(II\) directly from region \(I\).  Curiously, we shall see that this topological twist in the definition of the sign of the Hamiltonian, can be handled without any complications in our new theory.	

\def\makeline{\vskip100pt \noindent = = = = = = = = = = = = = = = = = =\\ }

As stated above in this Section, particles, or more precisely, the fields of the particles, must be defined both in region \(I\) and in region \(II\). All these particles must be physical; this is our first departure from older wisdom. This is important, because both regions \(I\) and \(II\) have their space- and time-like asymptotic domains at infinity.

Next, we distinguish particles going in, henceforth referred to as in-particles, from particles going out, the out-particles. In-particles cross the future event horizon (\(u^-=0\)), at a position given by the light cone coordinate \(u^+\), out-particles cross the past horizon (\(u^+=0\)) at a point given by \(u^-\). If \(u^\pm>0\), we are\fn{Note that sign conventions for light cone coordinates are often chosen differently.}  in region \(I\), if \(u^\pm<0\), we are in region \(II\). The distinction in- and out- makes sense as soon as particles move nearly with the speed of light in the longitudinal direction; the transverse velocity, and the mass, become negligible when they are sufficiently close to one of the horizons: close to a horizon, one may neglect \(\tl p\) and \(\m\), so that, according to Eq.~\eqn{massshell}, either \(p^+=0\) (an in-particle), or \(p^-=0\) (an out-particle).

Finally, we shall have to distinguish \emph{hard} particles and \emph{soft} particles. The distinction will be \emph{frame dependent} (we shall return to this point): a hard particle has its mass \(\m\) and/or momentum \(|\vecc p|\) of the order of, or beyond, the Planck mass. Soft particles have masses and momenta that are negligibly  affecting the curvature of the surrounding metric.

Soft particles will be given by their fields, including first time-derivatives where needed, on the Cauchy surface (dotted line in Fig.~\ref{penroses.fig}$a$). Their interactions will be described by whatever quantum field theory is applicable in energy domains close to the Planck scale, loosely indicated as `standard model interactions'. Their gravitational interactions need not be ignored, but, being weak, may be addressed in terms of \emph{perturbative} gravity. 

Hard particles are given by their geodesics. Again, we only need to consider them when they go almost with the speed of light in the longitudinal direction. The interaction between hard particles and soft particles also follows from standard theories: if the hard particles are charged, the effects of the charges on the soft particles are readily computed, and in this work we consider those interactions to be weak. Most important, however, is the gravitational interaction between hard particles and soft ones. The hard particles are hardly affected by the soft ones, but conversely, the soft particles feel the presence of the hard ones mainly through their gravitational forces. These are of the Shapiro type: soft particles are dragged along by the hard ones.\,\cite{Aichelb} This effect has been calculated and discussed in several of our previous publications,\cite{GtHSmatrix,GtHrecent1,GtHrecent2,GtHDray}. Most importantly, this effect diverges as the momenta of the hard particles increase.

We now make an important restriction on our description of black hole micro-states: we consider the Penrose diagram of Fig.\ \ref{penroses.fig}$a$, with \emph{only} soft particles added. Note that it is consistent then to ignore the diagrams of Fig.\ \ref{penroses.fig}$b$--$d$, because these contain hard particles (both the imploding matter and the Hawking particles there were assumed to affect the space-time curvature). We herewith insist that limiting ourselves to soft particles only, suffices to describe all black hole micro-states. What is meant by this will become clear when we deal with the `firewall transformations'. For now, we note that the soft particles in question are defined on the Cauchy surface shown; if they turn into hard particles elsewhere, this does not affect the state, but the effect must be considered when the evolution operators are studied.

Soft particles can become hard during the evolution, Eq.~\eqn{evolv}. The soft-hard interactions may have effects that are so large that soft particles may be moved around from region \(I\) to points deep inside region \(II\) and vice versa. This is why we cannot ignore region \(II\).
See also the discussion in Section~\ref{penrosemod}.

\begin{figure}[h!]
	\qqquad \includegraphics[width=320pt]{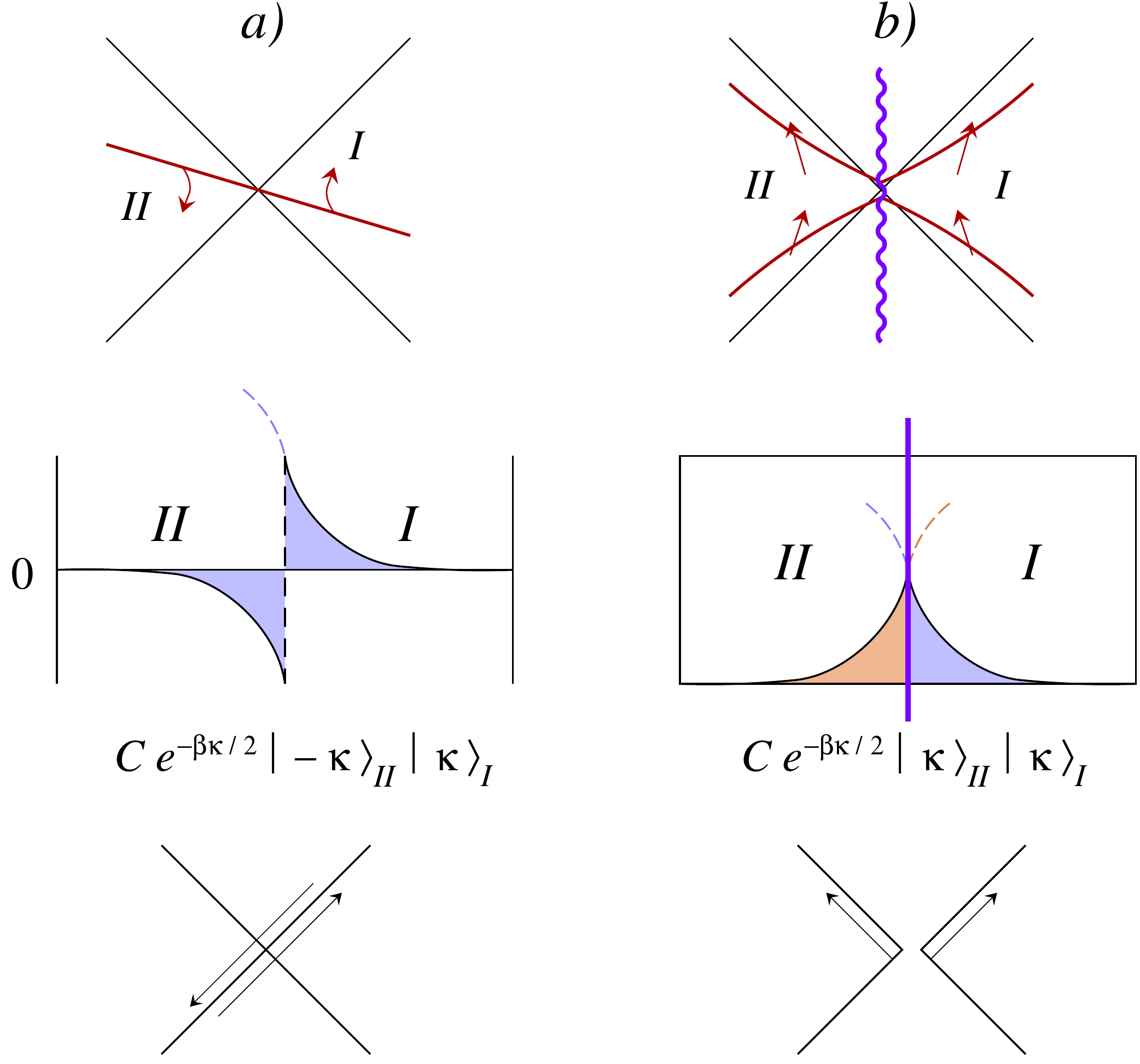} \begin{quote}
	\begin{caption}{ }{\small{Regions \(I\) and \(II\) of the Penrose diagram, $a)$ glued together locally smoothly, as seen by a local observer who regards region \(II\) as a continuation of region \(I\); $b)$ For a distant observer, the Hamiltonian density in region \(I\) and at the antipodes, represented by region \(II\), is positive. That observer would be tempted to time-reflect region \(II\).  $a)$ Above:  the Cauchy surface undergoes a Lorentz transformation, generated by \(L=H_I-H_{II}\), when there is a time boost for the distant observer.  Middle: according to local observers in region \(I\), the Cauchy surface in region \(II\) moves backwards in time, and therefore the energies of the quantum states there, are negative compared to the vacuum energy. The state shown here is close to the Hartle-Hawking state, which is the vacuum according to local inertial observers. Under \(b\), we see how this gives rise to positive energy Hawking particles both in \(I\) and in \(II\).  Below:  In the \(HH\) vacuum there are vacuum fluctuations. These have total energy = 0 according to \(a\), while the Hawking particles in \(I\) and \(II\) all have positive energies as seen by \(b\). }}	\labell{flippingregions.fig}
	\end{caption}\end{quote}
\end{figure}

New in our theory will be the  postulate that both regions \(I\) and \(II\) describe physically accessible parts of the same black hole  (by applying the antipodal identification),  so that pure quantum states of the local observer map onto pure quantum states of the black hole as seen by a distant observer. 

The dotted lines in Fig.~\ref{flippingregions.fig} illustrate the fact that one might consider the energy levels of region \(II\) as ``nearly full" rather than nearly vacuum. Note that this picture applies to \emph{soft particles only}.

\subsection{Hawking radiation I\labell{Hawk1}}
We end this chapter with rephrasing the standard physical features of Hawking radiation. As we shall see, the usual arguments pertain in particular to soft particles. What happens with hard particles will be exposed in Chapter\ \ref{Hawk2}.

Locality in the Standard Model allows one to distinguish the fields that live in \(I\) from the fields that live in \(II\). If one uses creation and annihilation operators to describe the quantum states in these regions, \emph{in terms of the fields in regions \(I\) and \(II\)}, one finds that the operators \(a\) and \(a^\dag\) normally used by a local observer, now mix the fields in region \(I\) with those in region \(II\). To unmix them, one 
hits upon the necessity to perform Bogolyubov transformations\,\cite{Bogolyubov, GtHSmatrix}. This means that the creation of a particle according to a local observer, may result in a superposition of a particle created and an (anti-)particle annihilated as seen by a distant observer.  Thus, the distant observer sees particles where a local observer sees none.

One finds that the vacuum state  \(|\varnothing\ket\) as seen by an inertial observer close to the horizons, for distant observers takes the form
\beq|\varnothing\ket&\iss C\sum_{\k,n}\, e^{-\half\b \k}\,|\k,n\ket_I\,|\k,n\ket_{II}\\
&\iss C\prod_\w \sum_{N_\w,\,n_\w}e^{-\half\b N_\w\,\w}|N_\w,n_\w\ket_I\,|N_\w,n_\w\ket_{II}\ , \eeql{entangled}
where \(\b\) is the inverse Hawking temperature, \(\k\) is the energy for the distant observer, \(n\) is any other type of quantum number, and \(C\) is a normalisation constant. \(N_\w\) is the number of particles with energy \(\w\). Since the generator of Lorentz transformations in the \(u^\pm\) direction (as seen by the local observer) is \(L=H_I-H_{II}\), this vacuum state, \(|\varnothing\ket\), is invariant under Lorentz transformations around the origin (\(L|\varnothing\ket =0\)).

Eventually, however, region \(I\) and region \(II\) must be regarded in unison, so as to assure that we are describing pure quantum states (duly entangled) only.  One can choose whether to apply this equation only to  the in-particles, the out-particles, or both.

The four special vacuum-like states a black hole can be in were aptly defined by Matt Visser\,\cite{Visser}: he distinguishes
\bi{\bal} The Hartle-Hawking\,\cite{HawkingPen} vacuum: the state described as a pure state according to Eq.~\ref{entangled}, that, according to a local, inertial observer, has both the in-particles and the out-particles in the vacuum state: nothing in, nothing out, but the distant observer has a bath of  particles going in and out. These particles look thermal at first sight, 
but in our formalism,  this is not an ordinary thermal bath: the particles in one hemisphere will be strongly entangled with those in the other hemisphere, so that, on total, we have a single pure state.
\itm{\bal} The Boulware\,\cite{Boulware} vacuum: there is a pure vacuum at infinity, hence no in-particles and no out-particles for the distant observer,
\itm{\bal} The Unruh vacuum: only the out-particles of a decaying black hole are seen, the in-particles are absent, and
\itm{\bal} The ``Vacuum cleaner vacuum": the in-particles are in a Hartle-Hawking state, the out-particles are absent.
\ei
In general, we consider these states \emph{only} for the soft particles.

As Hamiltonian for a \emph{local, distant} observer (an observer only looking at one hemisphere), one can use the generator of the Lorentz transformation that keeps the horizons in place: \(L=H_I-H_{II}\). We have 
		\be H_I=\int_{x^3>0}\dd^3\vec x\,x^3\HH(\vec x,0)\ ,\quad H_{II}=\int_{x^3<0}\dd^3\vec x\,|x^3|\,\HH(\vec x,0)\ ; 
		\quad x^3\equiv\fract 1{\sqrt 2}(u^++u^-)\ , \ee
where \(\HH(\vec x,0)\) is the Hamiltonian density at time \(t=0\). 

We have, in the absence of gravitational interactions,  \([H_I,\,H_{II}]=0\). 
The eigen states of \((H_I,\,H_{II})\) are the Boulware states (excitations from the Boulware vacuum using finite products of creation operators),
 \(|\k_1,n_1\ket_I\,|\k_2,n_2\ket_{II}\).  Here, \(\k_1\) refers to the total energy eigenvalue of \(H_I\), and \(\k_2\) is the energy eigenvalue of \(H_{II}\). However, the Hartle Hawking state, representing the vacuum for the local observer, has both \(H_I\) and \(H_{II}\) highly divergent; there are infinitely many Boulware particles queuing up near the horizon, but in such a way that \(H_I-H_{II}=0\) (due to Lorentz invariance, \(L=0\)).

The apparent ambiguity of the sign of \(H_{II}\) will later be seen to be due to the fact that, if we follow a closed curve from \(I\) to \(II\), the sign of the time coordinate flips, as in a M\"obius strip (see Chapter \ref{conc}); while the distant observer experiences a Hamiltonian that looks as \(H=H_I+H_{II}\), the observer who looks at the Hamiltonian from near the horizon only, sees \(H=L=H_I-H_{II}\).

As long as we look at soft particles, this local Hamiltonian is conserved in time. However, soft particles may evolve into hard particles, and these can cross the horizon, as we shall see. When we include the hard particles, we have to use \(H_I+H_{II}\), plus an interaction part, as our Hamiltonian.

\newsecl{The hidden asymptotic region at the horizon}{hiddenasympt}

Very near to the horizon, both \(H_I\) and \(H_{II}\)  have an infinite degeneracy of eigen states. This would generate an infinity in the number of micro-states, which has to be addressed.

To describe the states \(|\k,\,n\ket\) as seen by a distant observer of a black hole, where \(\k\) is the energy and \(n\) represents other quantum numbers, it is appropriate to use the tortoise coordinates, \(\r\) and \(\t\):
	\beq \r&=\half\log\Big(\frac r{2GM}-1\Big)+\frac r{4GM}&=&\ \half\log(x\,y)\\
		\t&=\frac{t}{4GM}&=&\ \half \log(x/y)\ ,\eeql{tortoise}
where \(x\) and \(y\) are the Kruskal-Szekeres coordinates, see Appendix\ \ref{KSantipodes}. The tortoise coordinates \eqn{tortoise} are useful because they reproduce the flat space-time coordinates asymptotically.
In these coordinates, a massless scalar wave packet \(\f(\r,\,\t)\)  obeys
	\be \pa_\t^2\f=\Big(\pa_\r^2+2(2GM/r)^2\pa_\r+(4GM/r)^2\,\ell(\ell+1)\,(1-2GM/r)\Big)\f\ . \ee
Note that, near the horizon, the angular momentum term (and possible mass terms) become insignificant. When the field \(\f\) is properly normalised\fn{When we write \(\f(r,t)=K(\r)\f(\r,\t)\), where, close to the horizon, \(K(\r)\ra e^\r\), the first derivative term can be replaced by a potential term. This is the usual way to restore hermiticity of the Hamiltonian, and with that, the restoration of the interpretation of \(|\f|^2\) as a probability distribution.}, the linear derivative term turns into a potential term of lesser significance, and we find that plane waves will propagate inwards and outwards with velocities less than 1 in the \((\r,\,\t)\) coordinates. Since \(\r\) goes all the way to \(-\,\infty\), we have infinite trains of plane waves going in and out, infinitesimally close to the horizon. This is our `hidden asymptotic region'.

In a first approximation, the states \(|\k,\,n\ket\) are obtained from the solutions of \(\f(\r,\t)=e^{-i\k\t}\f(\r,0)\), where \(\k=\sum_\w N_\w\,\w\), when each mode \(\w\) is occupied by \(N_\w\) particles. These states form the basis of states as experienced by the distant observer; they contain \(N=\sum_\w N_\w\) particles.

As in Section \ref{Hawk1}, one would now be tempted to use the Hartle-Hawking state \eqn{entangled}.

The fact that there are infinitely many modes at \(\r\ra+\infty\) is easy to understand; this is due to the infinite amount of space outside the black hole, so we can imagine a cut-off by imposing boundary conditions of a large box surrounding the hole. 

The limit \(\r\ra-\infty\) is a different one and it is important.\fn{Later, these two infinite domains will be found to be related. See end of Section  \ref{clones}.\label{noteinfty}} This hidden asymptotic region is easy to interpret, but often, incorrectly, ignored: there are infinitely many modes \(\w\) describing particles queuing up at the horizon. It takes them forever to pass the horizon, but they do not naturally reflect backwards. Imposing a cut-off there, requires the following discussion\,\cite{GtHBH}:

If we would keep all quantum states generated by the fields on the domain \hbox{\(-\infty<\r<0\)}, we would get a strictly infinite spectrum of micro-states for the black hole, which clashes with our physical expectation that the number of micro-states should be finite and should agree with the thermodynamics of Hawking radiation. Both the  in-particles and the out-particles give infinite numbers of micro-states near the horizon.

Equivalently, we can blame the infinity to the simple fact that the vacuum state \(|\varnothing\ket\) is invariant under the subgroup of    longitudinal Lorentz transformations, whereas, for the local observer, this subset of the Lorentz group is non compact; Lorentz transformations generate infinite sequences of states. 

To adapt this situation to what we expect physically, we again have to \emph{modify the theory.} In-particles too close to the horizon should, somehow, be replaced by out-particles, which must carry all information along. This could be achieved by erecting a `brick wall'\,\cite{GtHBH}, but that would be too drastic and too difficult to justify physically, while a better solution is at hand.

What we expect physically, is that there should be a boundary condition at some large, negative value of \(\r\), relating \emph{all} in-particles  to out-particles. This is the statement that the process of formation and evaporation of a black hole is controlled by a unitary scattering matrix \(S\)\,\cite{GtHSmatrix}. 

Note however, that not only standard theories are unable to determine the form of \(S\),  but even the existence of such a matrix is often ignored\,\cite{Hawking}.

A way to phrase the new situation is to observe that our hidden asymptotic region must be sealed off. We now describe a natural mechanism that was found to do exactly that.

Since \(\t\) runs backwards in region \(II\), states \(\psi\) as seen by a local observer are the normal product of ordinary states in region \(I\) and the complex conjugates of ordinary states in region \(II\). This is why our first idea\,\cite{GtHdensity} in 1984 was to propose that the two regions \(I\) and \(II\) of the Penrose diagram, together represent states \(|\psi\ket\,\bra\psi|\), or, elements of the density matrix seen by the outside observer. This idea, however, also led to difficulties: in Ref.~\cite{GtHrecent1}, we found that there are direct transitions between states in region \(I\) and states in region \(II\), which would correspond to direct transitions between bra- and ket-states, and as such clash with unitarity.

Our recent investigations\,\cite{GtHrecent2} led to a hybrid theory -- in a sense, we combine the bra-ket idea of 1984 with the notion   that the time evolution should be unitary, \emph{i.e.}, it should \emph{not} mix bra states with ket states. Our solution is not just a wild theory or model of what could be a vague idea, but rather the contrary: it follows by explicitly inspecting the equations after an expansion in spherical harmonics. In terms of the spherical harmonics, our Schr\"odinger equation decouples all \(\ell,\,m\) components from one another, so that the equations are simple and unique differential equations in one space- and one time variable, with explicit solutions, so that there can be no doubt about their correctness. The antipodal mapping that connects regions \(I\) and \(II\) is seen to be inevitable. Let us continue describing the physics.

Our explicit analysis invited us to introduce the notion of {hard} and {soft particles}.  The hard particles are inevitable because, due to the evolution law \eqn{evolv}, \(p^\pm\) can grow or shrink exponentially as \(\t\ra\pm\infty\). However, both the mass \(\m\) and the transverse momenta \(\tl p\) do not grow or shrink. If we decide from now on to keep (in natural units)
\be |\tl p|\ll M_\Pl\ ,\quad \m\ll M_\Pl\ ,\quad \ell\ll M_{BH}/M_\Pl\ , \ee
then the only hard particles to keep track of have either \(|p^-|>M_\Pl\) (the hard in-particles) or \(|p^+|>M_\Pl\) (the hard out-particles).

It just so happens that hard elementary particles have never been observed. In describing our micro-states, \emph{we shall assume that they always can be omitted}, but in order to keep the hard in- and out-particles out of the way,  ``firewall transformations" will be needed, as will be explained\,---\,the assumption will be verified.

The firewall transformations were in fact implicitly used to arrive at the unitary matrix  in Refs.\ \cite{GtHrecent2} (see also Appendix \ref{ER-EPR}), as it connects only soft particles. However, these particles are described in terms of their contributions to the energy momentum distribution \(p^\pm(\tht,\vv)\) or their contributions to the average position distribution \(u^\pm(\tht,\vv)\). What then remains to be done is to 
 map these data onto the states of Fock space in the Standard Model. This however will have to be left for future investigations.

\newsecl{The correct construction of the micro-states}{microstates}	 \setcounter{footnote}{0}

We shall only make use of conventional quantum field theories when addressing soft particles. To formulate a theory for the black hole micro-states, hard particles, embedded in a non-trivial, curved background space-time,  are to be eliminated as follows. First, we define the quantum states in terms of only soft particles roaming around in the metric of an `eternal' black hole. Then, we show how to modify the evolution laws such that the evolution operator remains unitary within this Hilbert space\fn{A Hilbert space that naturally must include all in- and out-particles in the black hole's vicinity, but far from the horizon.}, and subsequently we can consider evolution over long time scales. To achieve the latter, text book physics needs modification concerning the boundary conditions at the horizon---in various ways.

Eventually, our description is intended to include black holes formed by the implosion of matter, as well as the final explosion of a black hole. But, to begin with, we only include Hawking particles that are emitted during a time interval that is short compared to \(M\log M\) in Planck units, and soft particles entering the hole during the same short period. This may seem to \emph{exclude} black holes emitting Hawking particles at much later times, and also all  black holes with the history of an implosion at much earlier times. These will be included however, through the firewall transformation, see later (Section \ref{clones}).

Hard particles with large values of \(\tl p\) and/or their masses \(\m\), are strongly (that is, exponentially) suppressed in the Hawking radiation, and usually not assumed to be present in the in-states either. So as stated earlier, we ignore those.

At time scales much longer than \(M\log M\) in Planck units, the Hawking particles appear to generate firewalls: due to continued Lorentz contractions, the \(p^\pm\) of these particles would diverge so fast that their effects on the metric can no longer be ignored. These are the hard particles that will be considered shortly.

Thus, as yet, neither are we concerned about the ancient history of the black hole, nor about its distant future, and note that this is standard practice when describing more conventional quantum processes: as soon as we have the complete set of states, the evolution laws for short time intervals are all one needs to know, to uncover the full time evolution features, by repeatedly applying the evolution laws found.

Having thus (temporarily) eliminated back reactions on the metric, we can now safely employ the full Penrose diagram of a stationary black hole, the one shown in Figure \ref{penroses.fig}$a$. We fill it with soft particles only, so neither the imploding matter of the distant past, nor the Hawking particles emitted in the late future, are visible.

To repeat: this description of the quantum states can only last for short time intervals. If time \(\t\) (as defined in Eqs.~\eqn{tortoise}) is allowed to become too large, then the in-particles are seen to be boosted so much that they violate the no-back-reaction condition; similarly, if we look at the far past, we see that the later out-particles obtain too much energy there, so that they too, fall out of the allowed domain.

This is a problem, which can now be cured. Consider an operator  creating an in-going particle\fn{This is \emph{not} exactly the creation operator \(a^\dag(p)\), but an operator increasing the ingoing \emph{total} momentum by an amount \((p)\), while keeping the norm of the state unchanged; hence, it is a unitary operator.\labell{aadagger}} with a very modest amount of momentum \(p^-\). The quantity \(G\,p^\m\) is so small that its effect on the curvature may be ignored. Now, as time \(\t\) proceeds, the component \(p^-\) of the momentum increases as
	\be p^-(\t)=p^-(0)\,\ex{\t}\ , \ee
while the distance \(u^+\) from the past event horizon (defined at the moment when the particle crosses the future event horizon)  decreases as
	\be u^+(\t)=u^+(0)\,\ex{-\t}\ . \ee
Notice, that the uncertainty relation 
	\be \d u^+\,\d p^-\gtrsim \hbar\ ,\ee
is not affected by the time evolution.

We then arrive at the point where we can no longer ignore the back reaction. Consider any other particle, a (Hawking) particle, going out. Its distance \(u^-\) from the future event horizon is shifted by the in-going object, in a way that can be computed precisely. It is in fact the only component of the curvature caused by the \(p^-\) particle that we have to take into account.
The shift \(\d u^-(\tht,\vv)\) depends on the location \(\W=(\tht,\vv)\) on the horizon, and was derived to be\,\cite{GtHDray}:
	\beq \d u^-(\W) =8\pi G R^{-2}\,f(\W,\W')p^-&\ ,&\W=(\tht,\vv)\ ,&\qquad\quad\W'=(\tht',\vv')\ ; \\
	(1-\D_\W)f = \d^2(\W,\W')&\ ,&\qquad\D_\W\,Y_{\ell\,m}(\tht,\vv)&=-\ell(\ell+1)Y_{\ell\,m}(\tht,\vv)\ ,  \eeql{shift}
where \(R=2GM_{BH}\) and \(\W'\) denotes the point on the horizon where \(p^-\) enters.\fn{The equations depend on the units chosen.  Since we work with the tortoise coordinates \eqn{tortoise}, the  variables \(u^-\) and \(p^-\) are dimensionless. In units where \(\hbar\!=\!c\!=\!1\), \(G\) has dimension length-squared.} 

As described in Refs.\ \cite{GtHSmatrix,GtHrecent1,GtHrecent2}, we use the fact that these equations are linear in the momentum \(p^-\) of the in-going particle. Repeat the argument for all particles that have \(p^-\) growing to values that become too large. They form a momentum distribution 
	\be p^-(\W) =\sum_ip_i^-\d^2(\W,\,\W_i)\ , \eel{inmomentumdistr}
and the total shift of all out-going particle positions is
	\be\d u^-(\W)=8\pi G R^{-2} \int\dd^2\W'\,f(\W,\W')\,p^-(\W')\ . \eel{outpositiondistr}
Next, we realise that the contributions of infinitesimally small \(p^-\) values in a somewhat more distant past can simply be chosen such that they generate the original positions \(u^-\) of all out-going particles. This argument justifies the idea that we can simply replace
	\be \d u^-(\W)\ \ra\ u^-(\W)\ , \ee
or, the \emph{average positions} \(u^-(\W)\) of all particles leaving the black hole at any solid angle \(\W\), are directly given by the \emph{momentum distribution} \(p^-(\W')\) on the in-going ones.

And finally, it is tempting to perform a spherical wave expansion\,\cite{GtHrecent1,GtHrecent2}, to arrive at the algebra: \begin{subequations}\begin{align}
		u^\pm(\W)=\sum\limits_{\ell,m}u_{\ell m}Y_{\ell m}(\W)\ ,
		&& \qquad p^\pm(\W) =\sum\limits_{\ell,m}p^\pm_{\ell m}Y_{\ell m}(\W)\ ;& \\  
		 [u^\pm(\W),\,p^\mp(\W')]=i\d^2(\W,\,\W')\ ,&&	
		[u^\pm_{\ell m},\,p^\mp_{\ell' m'}] =i\d_{\ell \ell'}\d_{mm'}\ ;& \labell{ellmcomm}\\ 
	u^-_\outt=\frac{8\pi G/R^2} {\ell^2+\ell+1}p^-_\inn\ ,&& 
	 u^+_\inn=-\frac{8\pi G/R^2} {\ell^2+\ell+1}p^+_\outt\ . & \labell{uprelations}		
	 \end{align}\labell{algebra}\end{subequations}

Three more steps are needed to arrive at a description of micro-states in terms of soft particles only. First, we need to establish what exactly the operators \(p^\pm_{\ell,m}\) and \(u^\pm_{\ell,m}\) mean, in terms of Standard Model particles. They are the spherical wave expansions of the operators \(p^\pm(\tht,\vv)\) and \(u^\pm(\tht,\vv)\), which in turn describe the \emph{total-momentum distribution} and the \emph{average position operator} of in- and out-going particles across the event horizon (specified at each solid angle \(\W=(\tht,\vv)\) separately). In this quality, they neatly obey our algebraic commutator equation \eqn{ellmcomm}. This does mean that \(p^-\) can be interpreted as the \(T^{--}\) component of the energy momentum operator, integrated over \(u^+\), and \emph{mutatis mutandis} \(p^+\). For any quantum state in the Standard Model, we can compute these operators, just as we can compute\fn{Note that, in such calculations, the partial wave expansion is not a \emph{quantum} superposition of angular momentum states, but a linear decomposition of \emph{operators}. Thus, although our mathematics shows a strong resemblance to the hydrogen atom, the physical machinery described here is different in important ways.}  the average position operators \(u^\pm(\tht,\vv)\). 

However, the converse is more problematic. Our algebra dictates how the \(p^-\) distribution of the in-particles dictates the \(u^-\) distribution of the out-particles, and \emph{vice versa}. This only yields a unitary evolution law \emph{if the Standard Model states are uniquely described by these components of the energy momentum tensor}. 

This is not obvious, but in our earlier work\,\cite{GtHBHstrings} on the relation between the black hole states and strings, it was noted that our theory is geometrically related to string theory. Our operators \(p^\pm(\tht,\vv)\) have the same form as string vertex insertions; the horizon simply plays the role of a string world sheet. In string theory, the claim that the quantum states are described by string vertex insertions, is considered quite acceptable. We adopt the same verdict in the present formalism.\fn{In fact, our algebra is closely related to the Virasoro algebra used in string theory\,\cite{sustr}, apart from two subtleties: first, our surface is a Euclidean one while the string world sheet is Minkowskian, which means that we have ``strings" with an imaginary slope parameter \(\a'\), and secondly, the central charge is missing. The central charge shows up when the Lorentz group is made complete; in our theory, transverse Lorentz rotations have been disregarded, but they may well lead to the same complications as in string theory, necessitating central charges.}

This leaves open the question how one arrives at the desired Standard Model quantum state once \(T^{\m\n}\) is given. This is an important open question at this moment. Note that, sooner or later, we should observe that globally conserved quantum numbers such as baryon number, will be inadmissible in most black hole theories, and the question will be raised
how such constraints will arise in practice, and how they should be dealt with. 

All we can do at this stage is conjecture that the mapping from energy-momentum density operators to Standard Model Fock space states, will be a unitary one. This conjecture requires that, when enumerating the states, and when proper cut-offs are used, the total number of relevant Fock space states should be equal to the number of possible values for \(p^\pm\) and \(u^\pm\).

The second step taken in this chapter concerns the physical interpretation of regions \(I\) and \(II\). The gravitational shifts \(\d u^-(\tht,\vv)\) described in Eqs.~\eqn{shift}, can easily be seen to carry a particle over from region \(I\) to region \(II\), or back (when an amount of \(p^-\) is annihilated). Consequently, our algebra \eqn{algebra} only closes properly if all in-particles are allowed to superimpose states in \(I\) with states  in \(II\). The \(p^-\) states are Fourier transforms of the \(u^+\) states. These are sharply defined only if the in-particle position operators are allowed to have both signs; we should be allowed to have them enter in \(I\) or in \(II\) or in a superposition of such states; we cannot restrict ourselves to one sign of \(u^+\) only.
The same must be true for the out-particles. In Fig.~\ref{penroseinout.fig} the situation is illustrated. Both the \(u^\pm\) operators and the \(p^\pm\) operators must be allowed to range from positive to negative, in which case the different regions both get involved. So how can we ensure unitarity in our system of quantum states?

\begin{figure}[h!]
	\includegraphics[width=450pt]{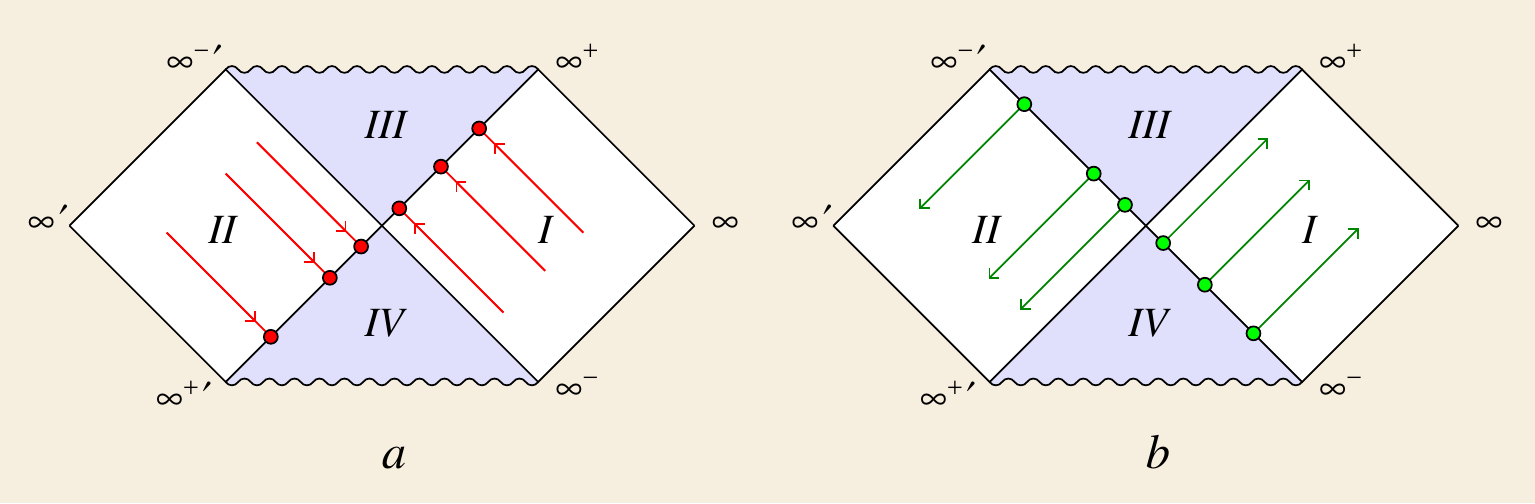} \\[-30pt] \begin{quote}
	\begin{caption}{ }{\small{The Penrose diagram with soft particles added. $a)$ Possible locations of the in-particles. At positive values of \(u^-\) they enter the hole in region \(I\), at negative \(u^-\) they enter the hole in region \(II\). $b)$ Showing possible out-particles. If \(u^+>0\), they emerge in region \(I\), if \(u^+<0\), they emerge in region \(II\).}}
	\labell{penroseinout.fig}
	\end{caption}\end{quote}
\end{figure}

The answer to this can only be that region \(I\) and region \(II\) must both represent physical black holes. Now, if these were allowed to be different black holes, then it would be inevitable to have cross-talk between these two, possibly widely separated, black holes, which would generate irreparable damage to any form of locality and causality.\fn{Often, it is brought forward that, therefore, these two black holes will be \emph{entangled}. This would not be a problem if the entanglement would be time-independent and hence not transmit information. In the present case, such entanglement would be influenced by in- and out-going material, and this would violate unitarity at the local scale. See Appendix \ref{ER-EPR}; finding such unitarity violation inadmissible, we searched for -- and found -- a better solution.}

It turns out that there exists exactly one clear and simple cure to this problem: \emph{regions  I and II represent the same black hole}.  Now if these would also be describing the same spot on the horizon, then this would generate conical singularities at the centre of the Penrose diagram. This would violate the principle of general relativity, since physics at a conical singularity is different from the physics of locally regular regions of space-time; we cannot allow this.

The correct answer must be that \(I\) and \(II\) represent different regions on the horizon of the same black hole. This means that we must have a \(\mathbb Z_2\) mapping of the horizon onto itself: if we move from one spot to its \(\mathbb Z_2\) image, we get the points connecting to region \(II\). The square of the mapping must be one, and there should be no fixed point. In Appendix \ref{ER-EPR}, we show that there is exactly one solution satisfying these constraints, which is that this \(\mathbb Z_2\) mapping is the antipodal mapping: \emph{moving from a fixed point of \((\tht,\vv)\) from \(I\) to \(II\) or back, must correspond to a transition to the antipodal points: }
	\be (\tht,\vv)\ \longleftrightarrow \ (\pi-\tht,\,\vv+\pi)\ . \eel{antipode}
Physically, such a space-time is remarkably regular. Since in all regions from which information can reach us, the radial coordinate \(r\) obeys \(r\ge 2GM_{BH}\), the points that we identify never come closer than \(2\pi GM_{BH}\), so local observers never notice anything unusual. In fact, one discovers two features that seem to be most welcome in a quantum theory of general relativity:
\bi{$i)$} Every point in our physical space-time now represents exactly one point in the Penrose diagram, not two, as we have in the conventional theory. Thus we uncovered a principle that may be a necessary one for the quantum theory: \emph{When fields on a manifold are quantised, it is essential that the entire asymptotic domain of the manifold maps one-to-one onto that of ordinary space-time, while preserving the metric.} (see the explanation of this statement in the Introduction, sect.~\ref{intro}: It must be possible to find time-like paths that connect all space-time points in \(\infty^+\)  to all points in \(\infty^-\)). This is the way to eliminate all problems with unitarity. 
% In particular, the difficulties with entanglement between late and early particles in the vicinity of the horizon, which prompted the authors of Ref.~\cite{firewall} to suspect the necessary presence of firewalls, are circumvented. Their firewalls, which indeed should have been a major concern, are transformed away.
\itm{$ii)$} All Cauchy surfaces must go through the origin of the Penrose diagram, so that, at a given time slice, \emph{there is no ``interior" region of the black hole.} As we never enter regions \(III\) or \(IV\) further than by an infinitesimal amount, all time slices used to describe black holes contain physically accessible points only (in the sense that they are connected to the outside world by time-like geodesics, both to the future and the past), with the exception of regions of measure zero. \ei
\begin{figure}[h!] \begin{center}
	\includegraphics[width=250pt]{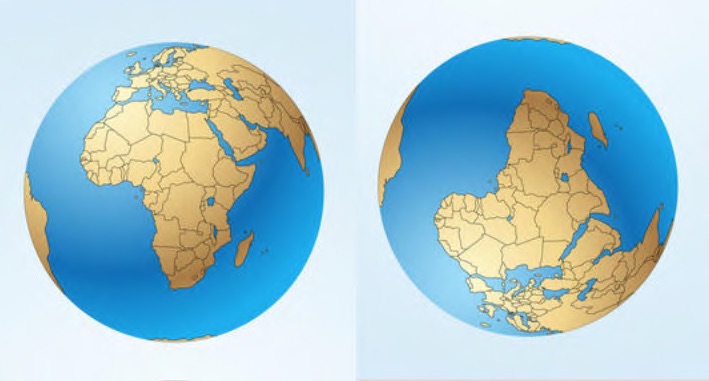} \end{center} \begin{quote}
	\begin{caption}{ }{\small{The antipodal mapping illustrated on planet Earth. Figure at the right shows the map formed by the antipodes of the figure at left, after rotating it \(180^\circ\) around the \(z\)-axis. Notice that the continents are  unfamiliar; they are parity-reflected.}}
	\labell{antipodes.fig}
	\end{caption}\end{quote}
\end{figure}

\subsection{Quantum clones\labell{clones}}

Now there is a third step to be taken. Looking at the in- and out-states of Figure~\ref{penroseinout.fig}, we see that, as time \(\t\) evolves, the in-states move further in, and the out-states move further out. This means that the momentum of the in-states increases exponentially as time runs forward, and for the out-states the momenta increase in the backwards time direction. In both time directions, therefore, we encounter hard particles (as defined in Chapters \ref{intro} and \ref{hiddenasympt}). The description of the micro-states as in Fig.~\ref{penroseinout.fig}, using soft particles only, therefore only works for time scales of the order of \(GM_{BH}\) in natural units.

If we want to cover larger time domains, our algebra \eqn{algebra} comes to the rescue. It states that, if in-momenta \(p^-\) become large, then the out-distances \(u^-\) will increase exponentially as well. Similarly, the \(p^+\) momenta of the out states follow the \(u^+\) positions of the in-states. Soon, the point will be reached that we are describing particles whose coordinates \(u^\pm(\tht,\vv)\) are so large that they have left the scenery of the black hole. \emph{Such particles may be ignored.} Note, that removing these particles that are far away now, allows us to redefine the black hole mass by subtracting the mass of the distant particles from the original expression. Thus, the black hole mass varies in time as expected physically.

Now comes something very important: the coordinates \(u^\pm\) are directly related to the momenta \(p^\pm\) according to Eqs.~\eqn{uprelations} in the algebra \eqn{algebra}. They do not represent different states, but refer to the same quantum states. What used to generate cloning problems now works as desired: the coordinates \(u^-\) and \(p^-\) are redundant, we only need one of the two to specify the quantum state. Similarly, we need either \(u^+\) or \(p^+\) to describe a single quantum state. To be able to ignore the gravitational back reaction, it is best to use these substitutions to obtain momenta \(p\)  when they are small, and/or coordinates \(u\) when they are big.

If we disregard particles that are too far separated from the black hole, we must, at the same time, disregard the high momentum particles to which they are linked according to Eqs.~\eqn{uprelations}. This is how we remove the particles that cease to be soft.

We can also phrase this is as follows: the hard in-particles, with values \(p^-(\W)\) for their momenta, are equally aptly described as if they were the soft out-particles, with \(p^-\) replaced by \(u^-\), in accordance with the algebra \eqn{algebra}. \emph{The hard in-particles are quantum clones of the soft out-particles}. The appropriate way to describe Hilbert space of all allowed states is to remove all particles with \(p^\pm\) exceeding some bound, by replacing them by the corresponding soft particles, for which \(u^{\pm}\) is now big enough to consider it as departed from the black hole. 

It was these ``hard" particles that generated ``firewalls" in earlier investigations\,\cite{firewall}. Now we see that there is a natural way to remove them. They are removed because they represent quantum states that can be better described as states containing soft particles sufficiently far away from the black hole's horizon.

Thus we replace all hard particles by soft ones, which also replaces in-particles by out-particles, and generates, in fact,  the scattering matrix relating in- to out-.  We refer to this transformation as the `firewall transformation'.  it is what we alluded to earlier: firewalls can be systematically and completely removed. Thus, our description of matter in terms of soft particles only, is validated a posteriori. As this includes the early matter particles that formed the black hole through an implosion, we see that now also our use of the eternal Penrose diagram, Fig.~\ref{penroses.fig}a, is justified. See also footnote (\ref{noteinfty}) in Chapter~\ref{hiddenasympt}: the hidden asymptotic region near the two horizons is linked to the physical asymptotic region infinitely far away from the black hole.

\newsecl{Novel aspects of this theory}{novel}	 \setcounter{footnote}{0}   	\def\maxx{{\mathrm{max}}}
	There is a number of important points that must be mentioned. We noted that the coordinates \(u^\pm(\tht,\vv)\), when positive, refer to region \(I\) and when negative refer to region \(II\), the antipode of \(I\). Something similar happens to the momenta. Upon careful examination of the algebra, we see that when a particle with momentum \(p^-\) enters the black hole in region \(I\), it has the same effect as a particle with momentum \(-p^-\) entering at the antipodal point. Here also, one must avoid double counting. If these two configurations lead to the same coordinate configuration for the out-going particles, then they must represent the same state. Thus, the momentum entering a point \((\tht,\vv)\) on the horizon always equals minus the momentum entering at the antipodal point. It is the demand that the \(u^\pm\) variables, by definition, have opposite signs in region \(I\) and region \(II\), from which we infer that, in the spherical wave expansion, \emph{only the odd values of \(\ell\) contribute}. 
	
This does not lead to any direct contradiction; physical particles can enter the black hole from both sides independently. This is because the energies \(\k\), as experienced by the outside observer, are something totally different from the momentum variables \(p^\pm\), which are the ones observed by the observer near the intersection of future and past event horizon. What one has to keep in mind is that, every in-going and out-going particle, enters or leaves at a \emph{different} point \((\tht,\vv)\) on the horizon. This is so because all in- and out-particles carry exactly one \(u^\pm\) coordinate, which is allowed either to be positive or negative, but not both. A particle with positive \(u^+\) enters in region \(I\), and negative \(u^+\) in region \(II\). The momenta \(p^-\) are then ill-defined because of the uncertainty relation. In the momentum frame, positive \(p^-\) means either a positive contribution in region \(I\) or a negative contribution in region \(II\). The energies \(\k\) are defined for regions \(I\) and \(II\) separately. At all \(\k\), the wave functions in regions \(I\) and \(II\) \emph{are} different, so that the energy distribution over the horizon is an arbitrary function of \(\tht\) and \(\vv\). The firewall transformation helps us to transform away states where either one  \(u\) coordinate is too small or one of the  momenta \(p\) is too large (both in Planck units).

Another important point is that the spherical wave expansion mixes up positive and negative signs (every \(Y_{\ell,m}\) function swings from positive to negative values), so as soon as we look at definite \((\ell,m)\) states, we mix up particles at one hemisphere with particles at the other hemisphere. 

The energies \(\k\) are the energies associated to the external time variable \(t\). In our dynamical equations, these energies are conserved regardless whether parts of a wave switch from region \(I\) to \(II\) or back. Thus, total energy will be conserved, in spite of the sign switch of \emph{local} energies discussed in Section~\ref{Hawk1}. 

The mode with \(\ell=m=0\) does not exist, as we just noted that \(\ell\) has to be odd. This implies that we cannot handle a single `dust shell' entering (or leaving) the black hole. What this really means is that such a `dust shell state' is ill-defined. The dust shell actually consists of myriads of `dust particles', each of which being allowed to be in a number of different quantum states. If a dust particle enters at one solid angle \((\tht,\vv)\), no particle is allowed to enter exactly at its antipode. It is perfectly allowed to enter at any other point very close by\,---\,but not too close\,\fn{The total number of points to be considered on the horizon will equal \(\ell_\maxx(\ell_\maxx+1)\), where \(\ell_\maxx\) is the cut-off for the spherical waves, as the transverse momenta there reach the Planck value. Since \(\ell_\maxx=\OO(M_{BH})\) in Planck units, this happens to be of the order of the total number of Hawking particles emitted by the average black hole\,\cite{Dvali}. \label{notelmax}}. \labell{pagelmax} 	
	
The energies of the in- and out-going particles, as observed by the external observer, refer to the plane waves in terms of the tortoise (Eddington-Finkelstein) coordinates \eqn{tortoise}. These tortoise coordinates commute with the sign operators \(\s^\pm\), since, close to the horizon, we have
	\be u^\pm=\s^\pm \ex{\r^\pm}\ . \eel{usigma}
Thus, at every point on the horizon\footnotemark[\value{footnote}] we will see just one particle entering or leaving, either at \((\tht,\vv)\) or at its antipode, but not at both.
 
The energy \(\k\) for each particle is independent of the local momenta \(p^\pm\), and can always be kept positive. Smearing the energies equally over all allowed solid angles gives us a dust shell, where ``dust" indeed stands for very many particles.
 \def\low#1{\raise-.3em\hbox{$\scriptstyle#1$}}

The Hamiltonian \(H_\t\) is the operator that causes the operators \(p^-\) and \(u^-\) to increase exponentially with \(\t\), while \(p^+\) and \(u^+\) should decrease exponentially. The operator that does this is the dilation operator. At each \(\ell\) and \(m\), we have (close to the horizon):

\beq H_\t\iss L\ &=&& H_I-H_{II}&&=&& \half(u^+\,p^-+p^-\,u^+)   		    &&=&& p^-\,u^++\half i    \\ 
	  &=&& i\big(p^-\frac{\pa}{\pa p^-}+\half\big) &&=&&  {-i}\big(u^+\frac{\pa}{\pa u^+}+\half\big)  &&=&&  {-i}\frac{\pa}{\pa\r} \ ,  \eeql{ham}
where the symmetrisation, which led to the terms \(\pm\half i\), was needed to keep the Hamiltonian hermitian. 

It was observed by Betzios, Gaddam and Papadoulaki\,\cite{drieling}, that this dilaton Hamiltonian in the variables \(u^\pm\) and \(p^\pm\) can be transformed into an apparently more conventional form, being the inverted harmonic oscillator. Rather than ellipses, the classical orbits in this potential are hyperbolas in phase space. An orbit may or may not bounce against the top of the potential, and on that it will depend whether a dynamical variable ends up in the same region or in the antipodal region of phase space.

 We keep our own notation. In a given spherical wave, energy eigenstates with energy \(\k\low{\ell m}\) have wave functions of the form
	\be \psi\ra C\,\ex{\,i\k_{\ell m}\,\r}\ . \eel{wavefunction}
In a quantum field theory, when disregarding the interactions, the Hamiltonians in regions \(I\) and \(II\) are\,\cite{GtHSmatrix} 
(Note the discussion on the sign of \(H_{II}\) at the end of Section~\ref{qustates}):
	\beq H_I&\iss\sum_i\int_0^\infty\dd\w\sum_{\ell,m} \w\,a^{I\,\dag}_{\ell m}(\w)\, a^I_{\ell m}(\w)\ ,&\\
	 H_{II}&\iss\sum_i\int_0^\infty\dd\w\sum_{\ell,m} \w\,a^{II\,\dag}_{\ell m}(\w)\, a^{II}_{\ell m}(\w)\ ,& \eeql{Hfields}
where the first summation symbol stands for the summation over different possible field types \(i\). The operators \(a_I\) and \(a_{II}\) are superpositions of the usual creation and annihilation operators of the field theory through a Bogolyubov transformation\,\cite{Bogolyubov}.

Observe that, in Eqs.\ \eqn{Hfields}, both \(H_I\) and \(H_{II}\) are non negative, while Eq.\ \eqn{ham} can have any sign. This has to be taken into account when mapping our states \(|\k_{\ell m}\ket\) onto the Standard Model states, a procedure that has not yet been elaborated to the author's satisfaction. 

One more remark here about the sign flip of \(H_{II}\). At first sight it looks as if the sign of this part of the Hamiltonian makes our theory inconsistent. Yet no problems were encountered in the explicit calculations. There is a good observation to be made to reassure us at this point. The crossing of the horizon, as described by the \(S\)-matrix also given in Appendix \ref{ER-EPR}, involves a substitution of the kind \(p\leftrightarrow u\), or, a Fourier transformation. As is well known, replacing \(p\) with \(u\) also replaces \(i\) with \(-i\). Now the matrix also contains terms on the diagonal, and these would generate the wrong sign. However, these terms are suppressed by factors \(e^{-\pi\k}\), so they cannot affect the signs. What counts, eventually, is that the matrix is unitary and the energy is conserved.

\subsection{The black hole interior\labell{interior}} The antipodal identification holds in particular for points situated on the horizon, \emph{i.e.}, the points at the centre of the Penrose diagram, Fig.~\ref{penroses.fig}\(a\). Outside the horizon, the points in region \(I\) are identified with the antipodal points of region \(II\). This means that the observable points outside the horizon (all living in region \(I\)), are not identified.

Nevertheless, the antipodal identification implies one important modification of our interpretation of the black hole metric, as compared to earlier work: \emph{the states \(|\k,\,n\ket_I\) and \(|\k,\,n\ket_{II}\) both refer to matter particles outside the horizon, so that the HH state \eqn{entangled} is a single, pure state.} This does away with the information problem and the unitarity problem in a radical fashion. Unitarity holds iff the regions \(I\) and \(II\) together represent the state the black hole is in.

Where then is the interior of the black hole? There is no interior. All equal-time lines (for the external observer) cross the centre point of the Penrose diagram, which is the intersection of future and past horizon, a space with measure zero. This means that, if one would move faster than the local velocity of light (that is, on a space-like geodesic), one could hop from a point on the horizon to its antipode instantly. It is as if, in 3-space, a sphere of radius \(R_{BH}=2GM_{BH}\) has been excavated, after which the antipodal points are glued together. In a conformal model, one could identify
	\be \vec x\equiv -\,R^2\vec x/|x|^2\ . \eel{excav}
Notice, that this effectively just removes all points \(\vec x\) with \(|\vec x|<R\). Notice also, that the transformation in \eqn{excav} does not invert the parity of local displacements \(\dd\vec x\).

In flat space-time, the transformation in Eq.~\eqn{excav} may seem to violate special relativity because two space-like separated points are identified. In the black hole, there is no local or global violation of special or general relativity. This we say because, on a time-like geodesic, the time it takes to reach a point at distance \(\e\) from the horizon, takes an amount of time that diverges as \(|\log\e|\). Once we reach the Planck scale, information can cross the horizon as it is spread over the Fourier transform of an out-going signal. Therefore, we can say that if we try to send information from some point \(r=a\) to its antipode, the trip through the horizon takes an amount of time of the order of \(R\,\log[(a-R)/R]\), while the detour around the black hole takes time of the order of \(\pi R\), where \(R=2GM_{BH}\) is the radius of the horizon. So the detour is actually faster. 

Even if our identification would suggest that one can beat the local velocity of light, this would still not have to violate special relativity because we have a preference frame induced by the background metric. We briefly continue on this subject in Appendix \ref{ER-EPR}.

The only `inside' region would be regions \(III\) and \(IV\). For the outside observer, however, these would be regions where his time coordinate is `beyond infinity', or `before minus infinity', and therefore these regions are unphysical for the outside observer.

A different -- but equivalent -- way of saying this is that regions \(III\) and \(IV\) of the Penrose diagram of the eternal black hole, contain just quantum clones of the particles in regions \(I\) and \(II\). A Cauchy surface must cover regions \(I\) and \(II\) \emph{or} be replaced partly to go through \(III\) or \(IV\), but care must be taken not to count any physically relevant degree of freedom more than once. This should be guaranteed if the Cauchy surface is space-like everywhere.

\subsection{The central singularity\labell{singular}}

Eventually, however, the antipodal identification does have an effect on the singularity structure of the black hole space-time. At all space-time points on, or in the immediate neighbourhood of, the horizon, antipodal points are all separated from one another by distances close to \(\pi R_{BH}\). Therefore, the identification is smooth and regular everywhere near the horizon.

Only at the plane \(r=0\), the effect of the antipodal identification will be more profound. The metric actually already has a singularity there, since, near the origin, its \(r\) component takes the shape \(-C\,\dd(r^{3/2})^2\) (being time-like in this region). The two branches here must again be identified antipodes, so that, perhaps, \(r^{3/2}\) is a more natural coordinate. Note however, that we have cosmic censorship in the Schwarzschild metric: the \(r\ra 0\) singularity occurs when, according to outside observers, time would be beyond infinity, so that no physical clashes occur at finite time. See footnote\ \eqn{censornote}, page \pageref{pagecensornote}.

\newsecl{Hawking radiation II}{Hawk2} % \setcounter{footnote}{0}

Although one might still express some doubts about the true nature of Hawking particles at opposite sides of the black hole, we have noted that, applying Eq.~\eqn{entangled} to describe them,  suggests a strong entanglement.  This then would lead to a prediction, if ever experiments could be done with radiating black holes. We now shall explain our prediction that, if at one point \((\tht,\,\vv)\) on the black hole horizon a Hawking particle is detected in spite of being suppressed by some fairly large Boltzmann factor \(e^{-\b E}\), then its anti-particle should emerge at the antipode without any further suppression.

The situation can be read off from the diagrams on the lowest line of Fig.~\ref{flippingregions.fig}\(a\) and \(b\). Fig.~\ref{flippingregions.fig}\(a\) is how one usually understands the emergence of Hawking radiation from local vacuum fluctuations at the horizon. A particle (upwards arrow) is created together with its antiparticle (arrow down). Any quantum numbers are arranged as following the arrows, so that the antiparticle has opposite quantum numbers -- it is a \(C\) inversion of the particle. When we flip it back \((b)\), we would be be tempted to make another \(C\) inversion, so that one might expect a particle at the point \(\vec x\) to be emitted together with the same particle at \(-\vec x\). This is what we expected in Ref.~\cite{GtHrecent2}. 

However, the points \(-\vec x\) are parity reflections of the point \(\vec x\). Indeed, if one would consider the map of planet Earth generated by the antipodes of the conventional map, one would notice it to be parity reflected. Quantum field theory of the elementary particles is not invariant under parity \(P\), but only under the combination \(CPT\), where \(C\) is charge conjugation and \(T\) is time reflection. Consequently, the parity image of a Hawking particle might not even exist (think of left-handed neutrinos). Thus, if parity is inverted, we expect the left-handed neutrinos to be replaced by right-handed ones. In addition, time must be reflected, but this is easy to interpret: the Hawking particles stay on one trajectory while the local time parameter flips from one sign to the other, so time reversal should not affect the particles.\fn{But this does mean that, at both sides of the horizon, we see the entangled particles at different times, as if \(t_2=-t_1\). Consequently, the particles will not behave identically. As soon as they left the horizon, one particle might decay differently from the other.}

We conclude that Hawking particles at opposite points of the black hole are each other's \(CPT\) reflections. Our conclusion at present is:\\
\emph{Hawking particles emerging from the black hole horizon are \(100\%\) entangled}: if a particle emerges that is strongly suppressed by the thermal Boltzmann factor  \(e^{-\vphantom{|^k}E/kT_{\scriptstyle{\mathrm{Hawking}}}}\), then at the antipode the antiparticle emerges with \(100\%\) probability, and in particular, with no further Boltzmann suppression at all. \emph{Locally}, we see a perfect thermal mixture with Hawking's expression for the temperature, but globally not: the two entangled particles together are suppressed by exactly one Boltzmann factor \(e^{-E/kT_{\scriptstyle{\mathrm{Hawking}}}}\), while in a thermal state one would have expected the square of that.

The subject of the entanglement of antipodal Hawking particles is still under investigation. It is also not yet clear how the actual value of the Hawking entropy can be deduced from the micro-states in this scheme. Locally, close to the horizon, we do expect Hawking's original value for the temperature to emerge, although, due to the entanglement, and our general philosophy that provides the black hole with pure quantum states, the entire black hole is not thermal at all.

\newsecl{Modifications of more conventional views; misconceptions and criticism}{mods}   %\setcounter{footnote}{0}
In discussions, the author became more aware of the thorny points in our arguments. Some of these we now discuss.

\subsection{The approximations made\labell{approx}}
Two simplifications were assumed, in order  to make our calculations possible. First, we ignored non-gravitational interactions, in particular electromagnetism. Actually, with some extra effort, electromagnetism can be included, but since the fine structure constant \(\a\) is not infinitesimally small, higher order corrections would become more problematic. Electromagnetism can be included by treating it as a gravitational force in a compactified 5\th dimension. It produces extra components to our algebra\,\cite{GtHSmatrix}. 

Primarily, this and other standard model interactions would have to take place during the short time interval when an out-going particle meets an in-going one. As long as these effects are perturbative and renormalizable, we expect them to be small. In any case, most other treatments of black holes also disregard such effects. A systematic study of standard model interactions at the black hole horizon will be interesting and important, but we expect the basic features discussed here not to undergo major changes due to these. 

A second approximation made was that we ignored gravitational interactions in the transverse direction: these cause in- and out going particles to shift sideways. As long as \(\ell\ll\ell_{\mathrm{max}}\), these corrections are also expected to be small, but as soon as \(\ell\) approaches its maximal value, of the order of \(M_{BH}\) in Planck units, this effect cannot be ignored; indeed, it is assumed to be responsible for the existence of an upper bound for \(\ell\). So, our approximations become less accurate for spherical waves approaching this limiting value.\fn{These waves would have to accommodate for details of Planckian dimensions; one might expect string theories to deal with them, in principle, but in our formalism,  the Planck domain of our theory has not yet been carefully examined.}

For lower values of \(\ell\), however, we expect our results to be quite accurate. It was claimed that our result was a ``merely classical approximation"\,\cite{mersini}, but then this would apply equally to the spherical wave expansion of the standard hydrogen atom, since also there, the photon, responsible for the \(e^2/r^2\) potential, is  a quantum object. As in the hydrogen atom, also the gravitational potential between in- and out-going matter can be treated as if it were classical. In any case, the operators arising from it are commuting, and that is what counts.

\subsection{The Penrose diagram\labell{penrosemod}} In many treatises about black holes, fundamental distinctions are made between the Penrose diagrams Fig.~\ref{penroses.fig}$a,\ b,$ and $c$, while \ref{penroses.fig}$d$ is rarely mentioned. Outside the past horizon, region \(I\) represents the surrounding universe; it is the same in all these diagrams. Regions  \(II,\ III,\) and \( IV\) are different. These differences come about because matter going in along the past event horizon, or emerging along the future event horizon, is either omitted or included in the picture.  The fact that both quantum mechanics and general relativity are symmetric under time reversal, while the purported solutions are not, was rarely regarded as an oddity. 

In most other branches of physics, it is customary that the laws of evolution only refer to the state the system is in at the same moment, not to its distant past history, nor to what is to be expected in the distant future, yet in black hole physics, often different laws are expected, depending on whether we are dealing with a black hole with a collapse in its past, or with an ``eternal black hole", or whether Hawking radiation in the distant future will get entangled one way or another. This cannot be right.

In our approach, it is crucial that the black hole carries no memory of how and when it came into being,  nor is it affected by whether or not it will evaporate or accumulate more mass in the distant future. We emphasise that we only wish to study the evolution laws for a black hole during relatively short periods of time, so short that neither matter going in, nor matter going out has had time to accumulate on the horizons;  we claim that the effects of these distant areas in time should be irrelevant. This makes sense when we realise that the complete Hilbert space is spanned by having only soft particles on the Cauchy surface (dotted line) in the Penrose diagram of Fig.~\ref{penroses.fig}$a$ at all times. As soon as the time evolution turns a soft particle into a hard one, the firewall transformation can be applied to replace it by its \emph{quantum clone}, a soft particle again.

We have seen that the momentum \(p^-(\tht,\vv)\) going in, is linked, one-to-one, with the positions of particles going out. Either the \(u\) coordinates, or the \(p\) operators, suffice to characterise the states, just as what we have for the positions \(x\) and the momenta \(p\) for ordinary quantum particles. If, in a given spherical harmonic wave configuration \((\ell,m)\), we Fourier transform the momentum wave operator of an in-particle, we get the {quantum clone} of the out-particle wave with the same harmonic wave numbers \((\ell,m)\).
A particle going in in region \(I\), lives on in region \(III\), while a particle entering in region \(II\) lives on in region \(IV\). Thus we can say that, if we work with the eternal metric Fig.~\ref{penroses.fig}$a$, regions \(III\) and \(IV\) are  {quantum clones} of regions \(I\) and \(II\). For the outside observer, no contradiction arises since that observer has no access to regions \(III\) and \(IV\). We can also say that regions \(III\) and \(IV\) are our universe at an exotic time \(t\) ``beyond \(\pm\)infinity".

It is very important to consider carefully how the gravitational back reaction should be represented in the Penrose diagram. In the diagram depicted in Fig.~\ref{penroses.fig}$b$, the particles contributing to the original collapse were taken to be classical; they strongly affect space-time curvature, allowing for the Minkowski geometry at epochs before the black hole was formed, but the Hawking particles, which all together represent as much energy as the imploding particles did, here seem to leave no trace. This is actually defendable: the Hartle-Hawking state is a \emph{single quantum state}, which, for a local observer falling in, is indistinguishable from the vacuum state -- hence, no gravitational effect. 

However, the Hawking particles can emerge in multitudes of modes, basically forming what looks like a thermal ensemble, so we do not want to represent them as a single mode, but in terms of as many quantum states as one can imagine coming out. These states cannot all be identified to the vacuum state, so the Penrose diagram must be something else as well. Hawking had proposed Fig~\ref{penroses.fig}$c$. But why this asymmetry under time reversal? This asymmetry is linked to the fact that not all information that went into the black hole was expected to come out. Most researchers today think that that is not evidently correct.

Our way to treat in-going and out-going matter, entirely symmetrically under time reversal, forces us to use more general wave functions for the out-going material, which means that the asymptotically far parts, as depicted in Fig~\ref{penroses.fig}$b$, $c$ and $d$, actually should form quantum superpositions, for which Penrose diagrams are not suited. This is why we advocate to stick to soft particles, added any way we like in Fig~\ref{penroses.fig}$a$, after which we leave it to the firewall transformation to continue our quantum states to (much) later or earlier time epochs.

\subsection{The black hole's global history as an instanton\labell{instanton}} Eventually, what one wishes to describe is the black hole's entire history in terms of wave equations, from initial collapse all the way to the final explosion. This is what was considered to be the black hole scattering matrix; the idea was that the black hole here acts as a virtual intermediate ``particle". When we consider the entire event as a tunnelling event, the picture may be seen as an instanton. The dominating parts of tunnelling amplitudes can be derived from classical equations in Euclidean space, which is why classical solutions in Euclidean space, in particular those that obey topologically non-trivial boundary conditions, are often considered with interest in particle physics. The instanton that would be of relevance for the black hole would be one where a region is excavated from a topologically trivial domain of Euclidean space, after which antipodal points on its boundary are identified.\cite{drieling} 

Wick rotating back to Minkowski space, we find that the black hole still behaves as an instanton with the same internal boundary condition. The antipodes in question are antipodes both in space and in time, although one has to keep in mind that, formally, time stands still at the horizon. This means that the firewall transformation procedure, linking the positions \(u\) of in- and out going particles, applies in one stroke to all particles entering and leaving the black hole during its entire life time. It is in the infinitesimal neighbourhoods of the horizons where we see that time is also inverted.

\subsection{The spherical dust shell\labell{shell}}
	As was stated at the beginning of Section~\ref{novel}, we only allow in- and out-going shells in the form of spherical harmonics with odd values for \(\ell\). This also raised objections. A black hole with perfect spherical symmetry should form if the collapse starts with a perfectly spherical collapsing shell, that is, only the wave with \(\ell=0\) is excited. What happens with such a black hole? Hawking's derivation of its perfectly thermal spectrum seems to be immaculate.

Hawking's result however is a statistical one. He could not derive any pure quantum state. This means that, regarding pure states, the collapsing shell must have myriads of pure states to choose from. This agrees with our general procedure. All odd values of \(\ell\), as long as \(\ell<\ell_{\mathrm{max}}\) (see footnote \ref{notelmax}, page~\pageref{pagelmax}), participate. Yet we can neither put all \(u^+_{\ell m}\) equal to zero, nor all \(u^-_{\ell m}\), since in these cases, \(p^-_{\ell m}\) or \(p^+_{\ell m}\) would tend to infinity; these would have to be removed by our firewall removing transformation. If, on the other hand, \(u^-_{\ell m}\) would be put equal to infinity, so that all \(p^+_{\ell m} \) vanish, then we would be dealing with particles far away from the black hole. These particles, no longer of physical interest, formally blur our pure quantum states; averaging over them might reproduce Hawking's thermal state, so again, there is no immediate conflict.

How the local momentum distributions \(p^\pm\) are distributed over the spherical waves, as opposed to the energy distribution \(\k(\tht,\vv)\), is further discussed in Section~\ref{novel}.

We conclude that the pure, spherical dust shell is unsuitable to serve as a model for a single quantum state of a black hole.

\subsection{Black hole formation through  collapse\labell{collapse}}
An objection sometimes brought forward is that our approach does not explain how a black hole forms by collapse. By time reversal symmetry, this complaint should also apply to the final stages of the black hole evaporation process. We seem to be focussing only on small changes taking place when a black hole captures or emits amounts of matter small compared to the total black hole mass.

Indeed, our theory is not completely finished. When a large amount of matter is captured or emitted, the total black hole mass before and after, should be different. How to accommodate for this was not yet studied in much detail in our theory, but we can make some general remarks.

As is well-known, in classical physics, black hole collapse is accompanied by a horizon opening up at one or several points in space-time. In the spherically symmetric case, one only has one such point. Behind that point, call it \(O  \), there is a small region where the horizon takes the shape of a future-directed light cone. Behind that  horizon, we have the region from which no signal can escape to infinity. In every respect, this region is to be regarded as a `region \(III\)' opening up. As stated in Section~\ref{penrosemod}, region \(III\) is a quantum clone of regions \(I\) and \(II\). This means that we should use tortoise coordinates surrounding the point \(O\). We can now consider a space-like sphere of infinitesimal size \(\dd r\), at a moment \(\dd t\) after the space-time point \(O\), where again we identify the antipodes. While more particles cross the horizon inwards, we continue to describe the evolution process as before, but because much more matter moves inwards that outwards, the black hole rapidly grows towards its final size and mass.

What we see here is, that we can apply our procedures from here on to describe an evolving quantum state. The actual process of growth was not yet fully described  in our procedures, but this is a complicated non-stationary background that requires further work. What we do see is that our principle of antipodal identification begins while the black hole still is infinitesimal in size, or more likely, it is of Planckian dimensions. This is one of the glimpses of our `new physics'; the only thing not understood about it today, is how it starts up in the Planckian domain, a domain that is still little understood as of this day. Once we have antipodal identification, this identification continues to apply for the entire lifetime of the black hole, and no further new physics is needed to understand that.

Two important remarks about black hole growth: the total number of partial wave modes \((\ell,m)\), is limited by a maximal value, \(\ell_{\mathrm{max}}\approx M_{BH}\) in Planck units. This means that there is approximately one particle (or wave) entering the horizon (or leaving) per unit surface area in Planck units.\,\cite{Dvali} We could turn this observation around and define the horizon area as being equal to the total number of \((\ell,\, m)\) modes, in units yet to be derived. As more spherical waves enter, or as more particles enter, the black hole then automatically grows.

Combine this observation with the fact that, at the intersection point of future and past event horizons, the outside observer's notion of time is not directly applicable. This point is eternal; it assembles all particles coming in and all particles going out during the entire lifetime of the black hole. During this long period, the black hole mass may have varied wildly; whenever an out going particle has moved out of the black hole's vicinity, one naturally deducts its mass-energy from the total mass of the system, to observe a new mass value for the hole itself, and the converse action is applied when particles enter. 

Thus, the firewall transformation comes with an adjustment of the black hole total mass value, and with that, with the actual horizon area defining the radius \(R\)   in the transverse directions, such as it occurs in our algebra, Eqs.~\eqn{shift} and \eqn{matrix}. 

Our second remark is that the increase or decrease of the black hole mass itself will be taken care of automatically when we realise that the black hole is actually made of the in- and out-going particles. After the firewall transformation, we see particles going out. They originated from spots infinitesimally close to the horizon, at which time the energies \(\k\) were definitely included in the black hole mass parameter \(M_{BH}\), and the total flux of its gravity field. As an out-going particle moves further out, the moment comes that it should no longer be considered as part of the black hole. As measured by the total flux of gravity, one must subtract the mass-energy \(\k\) of the out-particle from the black hole mass value: \(M_{BH}\Rightarrow M_{BH}-\k\). So from this kinematical point of view the total mass of the black hole will be taken care of automatically.

\newsecl{Conclusions}{conc}

What is agreed upon by many investigators, is that `new physics' is needed to resolve the information paradox and the quantum cloning problem, which led to the necessity to take `firewalls' seriously. We do claim to have arrived at `new physics', but, perhaps surprisingly, our approach does not require any modifications either in general relativity or in quantum mechanics. Our first `new physics' step is to add an \emph{extra condition} to constrain the allowed general coordinate transformations (see Section \ref{microstates}):
	\begin{quote} 	\emph{When fields on a manifold are quantised, it is essential that the entire asymptotic domain of the manifold maps one-to-one onto that of ordinary space-time, while preserving the metric.}\end{quote}
	This does not hold for the Kruskal-Szekeres coordinates (they are one-to-two), which is why unitarity seems to be violated there. The cure is simple, but leads to important new space-time features: an element of a \(\Bbb Z(2)\) subgroup of \(O(3)\) has to be chosen to identify pairs of points in the Kruskal-Szekeres coordinates. As to how the identification should take place, we have no choice: the only element of the \(O(3)\) space-time symmetry group that obeys our requirement of also keeping our coordinates singularity free, is the element \(-\Bbb I\) of \(O(3)\), yielding the antipodal identification (as explained in Appendix \ref{ER-EPR}). The fact that the arrow of time flips at the horizon -- and with that, the sign of the Hamiltonian -- seems to be no major obstacle. To understand what  is going on, the expansion in spherical harmonics is essential.

We had found that the firewall problem is a serious complication when one attempts to deduce the quantum properties of black holes from standard physical theories such as general relativity and quantum field theory. The problem could not be easily solved by declaring firewalls to be unphysical; as was noted by several authors, one would ``need new physics for that". Many authors leave it at that. However, the roots of a solution were already present in our early papers\,\cite{GtHBH,GtHSmatrix,GtHBHstrings}, though not very explicit. The fact that this remained unobserved makes me suspect that, although these papers were cited, they were not carefully read.\fn{They were certainly summarised incorrectly.}

A much more satisfactory answer was found more recently\,\cite{GtHrecent1,GtHrecent2}: if one expands the in-going and out-going matter in spherical harmonics, then at given \(\ell,m\), partial differential equations in one space- and one time dimension are obtained, which can be solved by the average undergraduate student.
One then notices the effects of gravitational back reaction much more explicitly: in-going waves are transmuted into out-going waves. The Hamiltonian near the horizon is simply the dilation operator, \(-\half(u\cdot p+p\cdot u)\), where \(u\) is the position of the in-going wave and \(p\) its momentum. This implies that \(u\) shrinks towards the horizon exponentially in time, while \(p\) blows up.

The gravitational back reaction links the in-going wave to the out-going wave, by \emph{interchanging momentum and position.} Thus, for the out-going wave, the Hamiltonian is \(+\half(u\cdot p+p\cdot u)\),  so that \(u\) blows up and \(p\) shrinks. The fact that \(u\) expands means that these waves quickly depart from the horizon, and subsequently from the black hole itself.

Thus our second `new physics' step amounts to identifying the position operators for the out-going wave with the momentum of the in-going one. Thus, the particle going out is a quantum clone of a particle going in. To avoid double counting, we must keep \emph{either} the in-going particle (in the from of a spherical wave) \emph{or} the out-going one, but not both. Replacing in-going  by out-going or \emph{vice versa} is what we call the `firewall transformation'. It results in a picture where all hard particles (in the longitudinal direction) can be replaced by soft ones.

Note that the spherical harmonics used here refer to operator distributions, not directly to wave functions (as in the hydrogen atom). The spherical harmonics refer to particle distributions, so surely all particles here are highly entangled, but since we have the explicit equations, there is no need to worry that `entanglement' would lead to any further problems here; all quantum states involved are in a well-defined basis of their Hilbert space, and how they evolve is uniquely determined, see the scattering matrix in Appendix \ref{ER-EPR}.

We emphasise that the wave functions we introduced on the variables \(u^\pm_{\ell,m}\) and/or \(p^\pm_{\ell,m}\) (obeying the usual commutation rules for positions and momenta) cannot and should not be `second quantised'. This is because \(u^\pm(\tht,\vv)\) already represent the \emph{average} positions of all particles at \((\tht,\vv)\), and \(p^\pm(\tht,\vv)\) represents the \emph{total} momentum there (as seen by a local observer close to the horizon).

Note also, that the antipodal identification of the points on the intersection of future and past horizon, is a crucial condition for the wave functions to remain pure.  We described the entanglement between Hawking particles emerging at antipodal points. This is arguably the most novel aspect of our `new physics'.

A mathematical curiosity is the fact that the antipodal identification comes with time inversion ($T$), besides parity ($P$) and charge conjugation ($C$). In space-time, consider a closed trajectory (not a geodesic) generated by a point that travels in outside space, first making a big  circle fragment from a point on the horizon to its antipode, then hopping back from the antipode to the point where it started. Projected on the horizon, the neighbourhood of this trajectory forms a M\"obius strip. Indeed, antipodal identification has turned the horizon into a (non orientable) projective plane, allowing M\"obius strips to be planted on it.\fn{Acknowledging a remark by D.G.\ Glynn, who also emphasises that  it should actually be called a "Listing strip" : ``Listing did more in the foundations of topology than M\"obius".} Curiously, our M\"obius strip is also time-like: going around it once causes the time coordinate to be inverted.

In other proposals for the resolution of the information paradox\,\cite{Hawkinginfo}, the points we noted were not observed, so that many mysteries were encountered that could not be resolved\,\cite{Polch}. 

On the other hand, we do not claim that all mysteries are resolved now. A systematic procedure must be found for a one-to-one mapping of the states generated by the spherical waves of momentum distributions and positions, onto states of the Fock space of a quantum field theory (some grand unified version of the standard model, relevant in the vicinity of the Planck scale, simply referred to as ``standard model" elsewhere in this paper). It is here that the machinery of string theory might be of much help.

An other point where our theory becomes vague is where the Planckian dimensions are reached. Usually it is assumed that string theory will provide all the answers, but string theory did not tell us about gravitational back reaction or antipodal identification, so we respectfully conclude that string theory is not fool-proof.\fn{Having said that, we do not wish to imply that string theory would be wrong; rather, we did not make use of it in our analysis. There is also the possibility that revised versions of string theory will enter: strings with a calculable, and purely imaginary, string tension parameter \(\a'\).}

\newsecl{Acknowledgments}{ack} \setcounter{footnote}{0}
The author  thanks  G.~Dvali, S.~Mukhanov, S,~Giddings, S.~Hawking, N.~Gaddam, O.~Papadoulaki, P.~Betzios,  A.~Franzen, F.~Scardigli, D.~Glynn, an unnamed referee, and many others, for discussions, questions and remarks.

\appendix
\newsecl{The antipodal identification for Kruskal-Szekeres coordinates}{KSantipodes}
The Kruskal-Szekeres coordinates \((x,\,y,\,\tht,\,\vv)\) are defined by
	\beq &x\,y&= &\ \left(\frac r{2GM}-1\right)\ex{r/(2GM)}\ , \\[3pt]
		&x/y&= &\ \  \ex{t/(2GM)}\ , \eeql{KScoord}
where \(r\) and \(t\) are the usual Schwarzschild coordinates, and \(\tht\) and \(\vv\) are unchanged. 
In terms of these coordinates, the usual Schwarzschild metric is found to be
	\beq &\dd s^2&=&\ -\dd t^2(1-\fract {2GM}r)+\frac{\dd r^2}{1-\fract{2GM}r}+r^2\dd\W^2& \\[3pt] 	
		&&=&\ \  \fract{32 G^3M^3}r\,\ex{-r/(2GM)}\,\dd x\,\dd y+r^2\dd\W^2\ ;&&\quad \dd\W^2\equiv\dd\tht^2+\sin^2\tht\,\dd\vv^2\ . \eeq
These coordinates are useful because they do not exhibit any manifest singularity at \(r=2GM\). 

The point emphasised in this paper is that every point \((r,\,t)\) in the Schwarzschild space-time is associated to \emph{two} points in the Kruskal-Szekeres notation: \((r,\,t)\) corresponds both with \((x,\,y)\) and \((-x,\,-y)\). This situation requires extra attention when quantised fields are considered in this metric. Usually, the points \((x,\,y,\,\tht,\,\vv)\) and  \((-x,\,-y,\,\tht,\,\vv)\) are assumed to be two different spots in space-time. When the space-time is disturbed by matter falling in or coming out, the points \((-x,\,-y,\,\tht,\,\vv)\) seem not to be there, but region \(III\) is there: \ \(x>0,\,y<0\), and it is tempting to extend space-time further. In our discussion, we consider states where soft particles roam about in a background not disturbed by matter at all, so it would include region \(II\) (the region \(x<0,\,x<0\)). This is a new representation of the black hole states, but we notice that it over-counts. To cure this shortcoming, we identify points, by the identification
	\be (-x,\,-y,\,\tht,\ \vv\,)\ \equiv\ (x,\,y,\,\pi-\tht,\,\vv+\pi)\ ,\eel{identify}
which we can also account for by stating that we constrain ourselves to (\(0\le\tht<\pi\), \(0\le\vv<\pi\)), imposing \eqn{identify} as a boundary condition.

{On} the intersection of the future and the past event horizon, \(x=y=0\), this turns the sphere into a projective sphere (a sphere with antipodal points identified), but when \((x,\,y)\ne (0,\,0)\), this is better to be seen as an identification of region \(II\) with region \(I\). Since far from the black hole we only have region \(I\), so this leaves the entire Schwarzschild space-time unchanged in the asymptotic domain; our identification just drops region \(II\). Thus, our identification poses no restriction on the states seen by an observer in the visible part of the universe; it does imply that, in this representation of black holes, the invisible part is absent, or equivalently, it is identified with the visible part.

In case black holes carry electric charges and/or angular momentum, the Reissner Nordstr\"om solution, the Kerr and the Newman et al solutions apply. These have also regions \(V\), \(VI\) and more, which do also contain asymptotic domains, but these regions lie either in the infinite future or the infinite past, so that they will play no role in our analysis. The Kerr and Newman solutions do require extra attention as the horizon rotates with super-luminal speeds, although we do not expect the need for major modifications of our analysis for these cases.

The singularity at \(r=0\) in the Schwarzschild metric and the inner horizon(s) in the Reissner-Nordstr\"om, Kerr and Newman solutions do become slightly more complex when our antipodal identification is applied, but this does not harm the theory because these singularities are shielded by the first horizon. They belong in regions where time goes beyond infinity or before minus infinity, where our theories need not be applied ever, 

We emphasise that, \emph{locally}, we do not deviate from that standard Schwarzschild metric,  its Kruskal-Szekeres extension, or its Penrose diagram, but \emph{globally} our modification is significant; it leads to big deviations from what was once thought to be a thermal state, and to new entanglements among the Hawking particles. Also we emphasise that our way of identifying antipodes does not lead to any singularity. This is because for all physical values of \(r\), its value is bounded by \(2GM-\e\) where \(\e\) is small. Singularities would only occur when \(r\) tends to zero. For the black hole, the region where this happens is unphysical; this would not have been the case if we assumed such an identification in flat space-time. In a polar coordinate representation of \emph{flat} space-time, this identification would indeed generate cusp-like singularities.

\newsecl{Locality violation for entangled black holes}{ER-EPR}
Some authors brought forward a slogan ``ER=EPR", meaning that the Einstein Rosen bridge yields entanglement as in the Einstein-Rosen-Podolsky set-up: the ER bridge is then assumed to connect different black holes. In the present theory, we claim that the ER bridge does not explain the phenomenon of entanglement, but it does lead to entanglement between the different black holes, to the extent that it causes a violation of locality, or local unitarity, unless the bridge connects a black hole with itself. Furthermore, general relativity and unitarity demand the absence of singularities on the horizon, so that we are only left with exactly one option: the ER bridge connects the antipodes on the horizon.

The violation of locality, in the case of different black holes, can be demonstrated explicitly. In  \emph{all} quantum field theories that were successful in the Standard Model, locality amounts to the demand that two localised operators \(\OO_1(x^{(1)})\) and    \(\OO_2(x^{(2)})\) must commute, \([\OO_1,\,\OO_2]=0\), if \(x^{(1)}\) and \(x^{(2)}\) are space-like separated points in space-time. Suppose that region \(II\) of a black hole would refer to another black hole (with exactly the same mass, charge, and angular momentum) that is space-like separated from the first one. Locality then demands that operators on black hole\,(1) should all commute with the operators on black hole\,(2), but we now show, that for regions \(I\) and \(II\), this is not the case.

Assume that the two black holes we start from, are not entangled, so we can call this state \(|\,I\,\ket\,|\,II\,\ket\). Now assume that we drop a particle at a position \(u^+\)  in black hole region \(I\). This we describe as the product state \(|\,I\,+ \{u^+\}_\inn\ket\,|\,II\,\ket\). It evolves into a state \(|\,I,\,II\ \,+\{p^+\}_\outt\ket\). This is a wave of particles going out in a state where it is a superposition of positions in black hole \(I\) and black hole \(II\). To get the position \(u^-\) for the out-particle, we have to Fourier transform the in-particle state \(\{u^+\}_\inn\), and this Fourier transform covers both regions \(I\) and \(II\), as was calculated explicitly in Ref\,\cite{GtHrecent2}. 

The scattering matrix \(S\)  for the energy eigen modes \(|\k\ket\) was derived in the spherical wave expansion:
	\be S\iss\begin{pmatrix} \a&\b\\ \b&\a\end{pmatrix}\iss
	\fract {e^{-\fractje {\pi i} 4}}{\sqrt{2\pi}}\G(\half-i\k) 		
		\left(\frac{8\pi G/R^2}{\ell^2+\ell+1}\right)^{\! -i\k} 
			\begin{pmatrix} e^{-\halfje\pi\k} &  ie^{+\halfje\pi\k}\, \cr  \,ie^{+\halfje\pi\k}& e^{-\halfje\pi\k} \end{pmatrix}\ ,
	\eel{matrix}
where \(R\) is the black hole radius, \(G\) is Newton's constant, \(\ell\) the partial wave coefficient, \(\k\) the wave number on the tortoise coordinates, also corresponding to the energy for the external observer; \(\G(x)\) is Euler's gamma function. 
The two elements of the wave function on which this matrix acts, correspond to regions \(I\) and \(II\) of Penrose diagram.  

One easily checks that unitarity holds, \(S\,S^\dag=\Bbb I\), only if we do include the off-diagonal elements of the matrix, which are indeed dominant if the energy \(\k>0\). One must conclude that, if the in-going particle enters exclusively in region \(I\), the out-going signal is a superposition of a signal going out in region \(I\) and a signal in region \(II\). Symbolically:
	\be |\inn\ket =|I+\{u^+_\inn\}\ket\,|II\ket\ ,\qquad |\outt\ket=\a|I+\{u^-_\outt\}\ket\,|II\ket\ +\ \b|I\ket\,|II+\{u^-_\outt\}\ket\ , \eel{inouttangle}
where \(|\a|^2+|\b|^2=1\ ,\ \ \b\ne 0\).		

Consider now the effects of an annihilation operator \(a_I\) acting on the added particle in region \(I\) and a similar operator \(a_{II}\) for region \(II\).
As we see in Eq.~\eqn{matrix}, the coefficient \(\a\) tends to zero if \(\k\) is large, so let us take the simplified case \(\a=0,\ \b=1\). In that case, 
consider the state \(|I\ket|II\ket\). We can create an in-going particle in region \(I\), after which an out-going particle in region \(II\) can be annihilated. Conversely, if we first annihilate the out-going particle in region \(II\), we get zero because the in-going particle was not yet created. Hence
	\be a_{II} \,a^\dag_{I}\,| I\ket|II\ket=|I\ket|II\ket\ ,\quad a^\dag_{I}\,a_{II}\,|I\ket|II\ket=0\quad\ra\quad [a_{II},\,a^\dag_I]\,|I\ket|II\ket\ne0\ . \ee
If regions \(I\) and \(II\) would refer to black holes that are far separated, in particular space-like, then the non-vanishing commutator assures that a signal can be sent. In conventional quantum field theories, space-like separation always guarantees that commutators vanish. This is why we maintain that regions \(I\) and \(II\) should not be taken to refer to widely separated black holes. 

What if they would refer to the same black hole? One could still ask, what if the two points considered are space-like separated? In that case, two observations can be made:
	\bi{$1.$} The two points are in the curved space-time background of a stationary black hole. If signals could be transmitted faster than light between these points, there would as yet be no clash with special or general relativity because we have a preference frame: the frame where the black hole is at rest. However,
	\itm{$2.$} The particle going in in region \(I\) needs a sizeable amount of time to reach the vicinity of the horizon, and after showing up in region \(II\), the emerging particle also requires a large amount of time to creep out. Both time lags are \(\OO(M_{BH}\log M_{BH})\) in Planck units, which goes to infinity in the classical limit. In practice therefore, these points are time-like separated.
	\ei
We conclude that locality forbids that regions \(I\) and \(II\) in Eq.~\eqn{matrix} refer to different black holes, but allows for the possibility that we are dealing with one and the same black hole, even if the points on the horizon are antipodal.

The  mapping must refer to different points on the horizon. Why are they \emph{antipodal} points? The answer to this is that the mapping 
must preserve the metric, so it is an element of \(O(3)\). Applying it twice must give the original point, so it is an element of a \(\mathbb Z_2\) subgroup of \(O(3)\). As an \(O(3) \) mapping, its eigenvalues are therefore \(\pm 1\). If any of its eigenvalues would be \(+1\), there would have been an invariant point, which would generate a conical singularity on the horizon. If we demand the absence of such singularities, we can only have all eigenvalues equal to \(-1\). This is the antipodal mapping.

One might observe that, actually, time is inverted as well (see Section~\ref{instanton}), so we have the element \(-\Bbb I\) of \(SO(3,1)\). It is still a special element in this group, but in Euclidean space, this element is homotopically identical to the identity itself. This is why any theory that can be analytically continued to Euclidean space, automatically obeys \(CPT\) symmetry. The antipodal identification is symmetric under \(CPT\).

%\makeline

\end{document}